\documentclass[aps,prd,10pt, twocolumn, superscriptaddress, floatfix, nofootinbib, amsmath, amssymb, preprintnumbers, longbibliography]{revtex4-1}

\usepackage{dcolumn}
\usepackage{verbatim}

\usepackage{breakurl}

\usepackage{lmodern}

\pdfoutput=1
\usepackage{graphicx}
\usepackage{bm}
\usepackage{amsmath}
\usepackage{amsfonts}
\usepackage{amssymb}
\usepackage{multirow}
\usepackage[colorlinks,linktocpage,linkcolor=cyan,citecolor=cyan]{hyperref}
\usepackage{adjustbox}
\usepackage{array}
\usepackage[capitalise]{cleveref}
\usepackage{acro}
\usepackage[utf8]{inputenc}
\usepackage{CJKutf8}
\usepackage{amssymb}
\usepackage{subfigure}

\usepackage{tikz}
\usepackage{pgfplots}
\pgfplotsset{compat=newest,every axis plot/.append style={line width=1pt}}

\crefname{figure}{Fig.}{Figs.}
\Crefname{figure}{Fig.}{Figs.}
\def\({\left(}
\def\){\right)}
\def\[{\left[}
\def\]{\right]}

\newcommand{\be}{{\begin{eqnarray}}}
	\newcommand{\ee}{{\end{eqnarray}}}

\newcommand{\Beq}{\begin{align}}
	\newcommand{\Eeq}{\end{align}}

\DeclareAcronym{SW}{
	short = SW ,
	long = Sachs-Wolfe ,
	short-plural =  ,
}
\DeclareAcronym{BH}{
	short = BH ,
	long = black hole ,
	short-plural = s ,
}
\DeclareAcronym{SNR}{
	short = SNR ,
	long = signal-to-noise ratio ,
	short-plural = s ,
}
\DeclareAcronym{IMRPPv2}{
	short = ,
	long = {\normalsize IMRP}{\footnotesize HENOM}{\normalsize P}v2 ,
	short-plural = ,
}
\DeclareAcronym{SFR}{
	short = SFR ,
	long = star formation rate ,
	short-plural =  ,
}
\DeclareAcronym{IMR}{
	short = IMR ,
	long = inspiral-merger-ringdown ,
	short-plural =  ,
}
\DeclareAcronym{ABH}{
	short = ABH ,
	long  = astrophysical black hole,
	short-plural = s ,
}
\DeclareAcronym{GW}{
	short = GW ,
	long = gravitational wave ,
	short-plural = s ,
}
\DeclareAcronym{SIGW}{
	short = SIGW ,
	long = scalar-induced gravitational wave ,
	short-plural = s ,
}
\DeclareAcronym{GWB}{
	short = GWB ,
	long = gravitational-wave background ,
	short-plural = s ,
}
\DeclareAcronym{CBC}{
	short = CBC ,
	long = compact binary coalescence ,
	short-plural = s ,
}
\DeclareAcronym{BBH}{
	short = BBH ,
	long = binary black hole ,
	short-plural = s ,
}
\DeclareAcronym{PBH}{
	short = PBH ,
	long = primordial black hole ,
	short-plural = s ,
}
\DeclareAcronym{LIGO}{
	short =LIGO ,
	long = Laser Interferometer Gravitational-Wave Observatory ,
	short-plural = ,
}
\DeclareAcronym{LVK}{
	short = LVK ,
	long = {Advanced LIGO, Virgo and KAGRA} ,
	short-plural = ,
}
\DeclareAcronym{ET}{
	short = ET ,
	long  = Einstein Telescope,
	short-plural =  ,
}
\DeclareAcronym{CE}{
	short = CE ,
	long  = Cosmic Explorer,
	short-plural =  ,
}
\DeclareAcronym{LISA}{
	short = LISA ,
	long  = Laser Interferometer Space Antenna,
	short-plural =  ,
}
\DeclareAcronym{BBO}{
	short = BBO ,
	long  = big bang observer,
	short-plural =  ,
}
\DeclareAcronym{DECIGO}{
	short = DECIGO ,
	long  = Deci-hertz Interferometer Gravitational wave Observatory,
	short-plural =  ,
}
\DeclareAcronym{SKA}{
	short = SKA ,
	long  = Square Kilometre Array,
	short-plural =  ,
}
\DeclareAcronym{PTA}{
	short = PTA ,
	long = pulsar timing array ,
	short-plural = s ,
}
\DeclareAcronym{FRW}{
	short = FRW ,
	long = Friedman-Robertson-Walker ,
	short-plural =  ,
}
\DeclareAcronym{CMB}{
	short = CMB ,
	long = cosmic microwave background ,
	short-plural =  ,
}
\DeclareAcronym{RD}{
	short = RD,
	long  = radiation-dominated ,
	short-plural =  ,
}
\DeclareAcronym{MD}{
	short = MD,
	long  = matter-dominated ,
	short-plural =  ,
}
\DeclareAcronym{HD}{
	short = HD,
	long  = Hellings-Downs ,
	short-plural =  ,
}
\DeclareAcronym{SMBH}{
	short = SMBH ,
	long  = super-massive black hole ,
	short-plural = s ,
}
\DeclareAcronym{SGWB}{
	short = SGWB ,
	long  = stochastic gravitational-wave background ,
	short-plural = s ,
}
\DeclareAcronym{NG15}{
	short = NG15 ,
	long  = NANOGrav 15-year ,
	short-plural =  ,
}
\DeclareAcronym{PSD}{
	short = PSD ,
	long  = power spectral density ,
	short-plural = s ,
}
\DeclareAcronym{PDF}{
	short = PDF ,
	long  = probability distribution function ,
	short-plural = s ,
}
\DeclareAcronym{BBN}{
	short = BBN ,
	long  = big-bang nucleosynthesis ,
	short-plural =  ,
}

\begin{document}

	\title{Constraints On Scalar-Induced Gravitational Waves Up To Third Order From Joint Analysis of BBN, CMB, And PTA Data}

	\author{Sai Wang}
	\email{Corresponding author: wangsai@ihep.ac.cn}
	\affiliation{Theoretical Physics Division, Institute of High Energy Physics, Chinese Academy of Sciences, Beijing 100049, China}
	\affiliation{University of Chinese Academy of Sciences, Beijing 100049, China}
	
	\author{Zhi-Chao Zhao}
	\affiliation{Department of Applied Physics, College of Science, China Agricultural University,
		Qinghua East Road, Beijing 100083, China}

		\author{Qing-Hua Zhu}
	\affiliation{CAS Key Laboratory of Theoretical Physics,
  Institute of Theoretical Physics, Chinese Academy of Sciences,
  Beijing 100190, China}
\affiliation{University of Chinese Academy of Sciences, Beijing 100049, China}

\begin{abstract}

Recently, strong evidence for a gravitational wave background has been reported by collaborations of pulsar timing arrays (PTA). In the framework of scalar-induced gravitational waves (SIGWs), we concurrently investigate the second and third order gravitational waves by jointly analyzing PTA data, alongside big-bang nucleosynthesis (BBN), and cosmic microwave background (CMB) datasets. We determine the primordial curvature spectral amplitude as $0.021<A_\zeta<0.085$ and the spectral peak frequency as $10^{-7.3}\ \mathrm{Hz}<f_\ast<10^{-6.3}\ \mathrm{Hz}$ at a 95\% confidence interval, pointing towards a mass range for primordial black holes of $10^{-4.5}M_\odot<m_{\mathrm{PBH}}<10^{-2.5}M_\odot$. Our findings suggest that third order gravitational waves contribute more significantly to the integrated energy density than the second order ones when $A_\zeta\gtrsim0.06$. Furthermore, we expect future PTA projects to validate these findings and provide robust means to investigate the genesis and evolution of the universe, especially inflation. 
\end{abstract}
	
	\maketitle
	
	\acresetall
	
{\emph{Introduction.}}
Theory of \acp{SIGW} \cite{Ananda:2006af,Baumann:2007zm,Mollerach:2003nq,Assadullahi:2009jc,Espinosa:2018eve,Kohri:2018awv} has been proposed to interpret the substantial evidence of nano-Hertz \ac{GWB} signals reported by several \ac{PTA} collaborations \cite{Xu:2023wog,Antoniadis:2023ott,NANOGrav:2023gor,Reardon:2023gzh}. It has been suggested to offer a better fit than the astrophysical interpretation in the context of \ac{SMBH} binaries \cite{Antoniadis:2023xlr,NANOGrav:2023hvm,NANOGrav:2023hfp}. Subsequent related studies are detailed in Refs.~\cite{Franciolini:2023pbf,Inomata:2023zup,Wang:2023ost,Liu:2023ymk,Abe:2023yrw,Ebadi:2023xhq,Figueroa:2023zhu,Yi:2023mbm,Madge:2023cak,Cai:2023dls}. Significant constraints have been imposed on the enhanced primordial curvature power spectrum at small scales, which are unreachable for conventional probes, such as the \ac{CMB} that is sensitive to physics on the largest observable universe scales \cite{Maggiore:2018sht}. However, these studies have solely considered second order gravitational waves, disregarding contributions from higher order ones. 

In addition to the direct measurements of \acp{SIGW} from the \ac{PTA} probe, early-universe probes offer indirect constraints on the integrated \ac{SIGW} spectrum \cite{Moore:2021ibq}. Given that \acp{SIGW} behave like radiation in the universe, their energy can alter the growth of cosmological density perturbations and the universe's expansion rate at the time of decoupling. Consequently, the \ac{CMB} probe is sensitive to the integrated \ac{SIGW} spectrum \cite{Smith:2006nka,Clarke:2020bil}. Simultaneously, the success of \ac{BBN} theory can limit the number of relativistic species at the nucleosynthesis epoch. Therefore, the \ac{BBN} probe is sensitive to the energy of \acp{SIGW} \cite{Cooke:2013cba}. Both probes independently measure cosmological \acp{GWB} of frequency bands above $10^{-10}$ Hz, but are insensitive to other \acp{GWB} produced due to astrophysical processes in the late universe.

In this study, we simultaneously consider second and third order gravitational waves, and explore joint constraint on them from \ac{BBN}, \ac{CMB}, and \ac{PTA} datasets. Previous research on third order gravitational waves is documented in Refs.~\cite{Yuan:2019udt,Zhou:2021vcw}. We will demonstrate that they dominate the \acp{SIGW}' energy density if the primordial curvature spectral amplitude exceeds $\mathcal{O}(0.06)$. We will also illustrate that they do not significantly alter the PTA bound but cause substantial changes in the \ac{BBN} and \ac{CMB} bounds. Consequently, the joint constraint will also experience significant alterations. Moreover, we will explore these possibilities and potential future improvements. 

 
{\emph{Scalar-Induced Gravitational Waves.}}
Adhering to the conventions of Ref.~\cite{Chang:2020tji}, we adopt the perturbed spatially-flat \acl{FRW} metric in Newtonian gauge, specifically,
\begin{eqnarray}
\mathrm{d}s^{2}&=&a^{2}\Big\{-(1+2 \phi^{(1)}+ \phi^{(2)}) \mathrm{d} \eta^{2}+ V_i^{(2)} \mathrm{d} \eta \mathrm{d} x^{i}\\&&+\big[(1-2 \psi^{(1)} -\psi^{(2)}) \delta_{i j}+\frac{1}{2} h_{i j}^{(2)}+\frac{1}{6} h_{i j}^{(3)}\big]\mathrm{d} x^{i} \mathrm{d} x^{j}\Big\} ~ ,\label{1}\nonumber
\end{eqnarray}
where the superscript $^{(n)}$ signifies the $n$-th order perturbations, $\phi$ and $\psi$ represent scalar perturbations, $V_i$ indicates transverse vector perturbations, and $h_{ij}$ denotes transverse-traceless tensor perturbations.

Tensor perturbations $h_{ij}$ induced by the linear scalar perturbations are referred to as \acp{SIGW}. 
Second order gravitational waves have been investigated in the literature \cite{Ananda:2006af,Baumann:2007zm,Mollerach:2003nq,Assadullahi:2009jc,Espinosa:2018eve,Kohri:2018awv}, and we follow the conventions of Ref.~\cite{Kohri:2018awv}. 
As shown in Appendix~\ref{A}, the equation of motion for third order gravitational waves is \cite{Zhou:2021vcw}
\begin{equation}
		h_{i j}^{(3)''}+2 \mathcal{H}  h_{i j}^{(3)'}-\Delta h_{i j}^{(3)}=-12 \Lambda_{i j}^{l m} \mathcal{S}^{(3)}_{l m} ~ ,\label{4}
\end{equation}
where $\mathcal{H}$ is conformal Hubble parameter, $\Delta$ is Laplacian, $\Lambda_{ij}^{lm}$ is transverse-traceless operator, and $\mathcal{S}_{lm}^{(3)}(\eta,\bm{x})$ denotes source terms, as expressed in Eq.~(\ref{5}). 
Leveraging the two-point correlators of $h_{ij, \bm k}^{(3)}$, i.e., $\langle h^{(3)}_{ij,\bm k} h^{(3)}_{ij,\bm k'} \rangle = 2(2\pi)^3 \delta(\bm k + \bm k') (2\pi^2)k^{-3} \mathcal{P}^{(3)}_h(k)$, we get the power spectrum for third order gravitational waves, i.e., 
\begin{widetext}
    \begin{eqnarray}
\mathcal{P}^{(3)}_h(k,\eta) & =& \frac{k^3}{32 \pi^2} \left(\frac{4}{9}\right)^3\sum_{\ast,\ast\ast}\int \frac{{\rm d}^3 p {\rm d}^3 q}{\left| \bm{k} - \bm{p} \right|^3 \left| \bm{p} - \bm{q} \right|^3 |\bm{q}|^3} \bigg\{ \mathcal{P}_{\zeta} \left( \left| \bm{k} - \bm{p} \right| \right) \mathcal{P}_{\zeta} \left( \left| \bm{p} - \bm{q} \right| \right) \mathcal{P}_{\zeta} (q) \nonumber\\
& & \times \mathbb{P}_{\ast,ij}\left( \bm k, \bm p,\bm q \right) I^{(3)}_{\ast} \left( | \bm{k} - \bm{p} |, | \bm{p} - \bm{q} |, | \bm{q} |, | \bm{p} |, k, \eta \right) \Big[\mathbb{P}_{\ast\ast,ij}\left( -\bm k, -\bm p,- \bm q \right) I^{(3)}_{\ast\ast} \left( | \bm{k} - \bm{p} |, | \bm{p} - \bm{q} |, | \bm{q} |, | \bm{p} |, k, \eta \right)\nonumber\\
 && +(\bm p \rightarrow \bm p - \bm q) +(\bm p \rightarrow \bm k - \bm q, \bm q \rightarrow \bm k - \bm p) +(\bm p \rightarrow \bm k - \bm q, \bm q \rightarrow \bm p - \bm q) \nonumber\\ &&+(\bm p \rightarrow \bm k - \bm p +\bm q)+(\bm p \rightarrow \bm k - \bm p +\bm q, \bm q \rightarrow \bm k - \bm p) \Big] \bigg\} ~, \label{13} 
\end{eqnarray}
\end{widetext}
where we introduce the quantities $\mathbb{P}_*(\bm k,\bm p, \bm q)$ and the kernel functions $I^{(3)}_*\left( p_1, p_2, p_3, p_4, k, \eta \right)$ in Appendix~\ref{A},  the subscripts $_{\ast}$ and $_{\ast\ast}$ denoting different sources of third order gravitational waves, i.e., in terms of $(\phi^{(1)})^3$, $\phi^{(1)} \psi^{(2)}$, $\phi^{(1)} V_i^{(2)}$, $\phi^{(1)} h_{ij}^{(2)}$.



Regarding the \ac{PBH} formation, there should be a large-amplitude peak on the power spectrum of primordial curvature perturbations \cite{Ivanov:1994pa,Garcia-Bellido:1996mdl,Yokoyama:1998pt}. Inflation models with sound speed resonance can generate a nearly monochromatic spectrum \cite{Cai:2018tuh,Cai:2019jah,Cai:2019bmk,Chen:2019zza,Chen:2020uhe,Yu:2023jrs}. 
For simplicity, we consider a delta-function spectrum 
\begin{equation}
\begin{aligned}
\mathcal{P}_{\zeta}(k)=A_\zeta k_{\ast}\delta(k-k_{\ast})~,
\end{aligned}\label{14}
\end{equation}
where $A_\zeta$ is the amplitude and $k_\ast$ is the pivot wavenumber. 
The energy-density fraction spectrum of \acp{SIGW} is defined as $\Omega_{\rm GW}(k,\eta)=2\pi G\langle \partial_l h_{ij}\partial_l h_{ij}\rangle/[{a^2\rho_{\rm c}(\eta)}]$ \cite{Kohri:2018awv}, where $h_{ij}\equiv(1/2)h_{ij}^{(2)}+(1/6)h_{ij}^{(3)}$, and $\rho_{\rm c}$ is critical density at conformal time $\eta$. 
We determine it as  
\begin{equation}
\Omega_{\rm GW}(k,\eta) = \frac{A^2_\zeta}{24}\left(\frac{k}{\mathcal{H}}\right)^2 \left[\mathcal{P}_h^{(2)}(k,\eta)+ \frac{A_\zeta}{9} \mathcal{P}_h^{(3)}(k,\eta)\right]~,
\end{equation}
where the power spectrum of second order gravitational waves, $\mathcal{P}_{h}^{(2)}(k,\eta)$, is calculated in pioneers' works \cite{Ananda:2006af,Baumann:2007zm,Mollerach:2003nq,Assadullahi:2009jc,Espinosa:2018eve,Kohri:2018awv}, and the power spectrum of third order ones, $\mathcal{P}_{h}^{(3)}(k,\eta)$, is shown in Eq.~(\ref{13}). 
Since gravitational waves behave like radiations, the energy-density fraction spectrum in the present universe is \cite{Wang:2019kaf}
 \begin{equation} 
     \Omega_{\mathrm{GW},0}(k) = \Omega_{\mathrm{r},0}
     \left[\frac{g_{\ast,\rho}(T)}{g_{\ast,\rho}(T_{\mathrm{eq}})}\right]\left[\frac{g_{\ast,s}(T_{\mathrm{eq}})}{g_{\ast,s}(T)}\right]^{\frac{4}{3}}
     \Omega_{\mathrm{GW}}(k,\eta) \ ,
 \end{equation}
where the physical energy-density fraction of radiations in the present universe is $\Omega_{\mathrm{r},0}h^{2} \simeq 4.2\times10^{-5}$ with $h=0.6766$ being the dimensionless Hubble constant \cite{Planck:2018vyg}, a subscript $_\mathrm{eq}$ denotes cosmological quantities at the epoch of matter-radiation equality, and both $g_{\ast,\rho}$ and $g_{\ast,s}$ stand for the effective relativistic specifies in the universe \cite{Saikawa:2018rcs}. 
Moreover, cosmic temperature $T$ is related with $k$, i.e., 
\begin{equation}
\frac{k}{\mathrm{nHz}} = 83.25 \left(\frac{T}{\mathrm{GeV}}\right) \left[\frac{g_{\ast,\rho}(T)}{106.75}\right]^{\frac{1}{2}}\left[\frac{g_{\ast,s}(T)}{106.75}\right]^{-\frac{1}{3}}\ .
\end{equation}

It should be emphasized that the contributions from the third order gravitational waves become significant on the small scales $k \gg k_{\ast}$. This is primarily due to the additional enhancement in their power spectrum originated from resonance of the higher order perturbations at late times. On the other hand, for the large scales $k \ll k_{\ast}$, the contributions from the third order gravitational waves can be neglected because their power spectrum is highly suppressed compared to that of the second order gravitational waves.

\begin{figure}
    \includegraphics[width =1 \columnwidth]{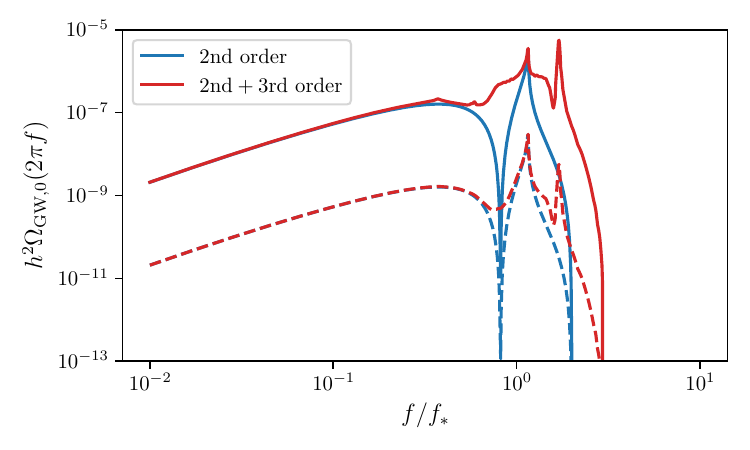}
    \caption{This figure contrasts the energy density spectra of second order (blue) with those of both second and third orders (red). The model parameters provided are $A_\zeta=0.1$ (solid) and $A_\zeta=0.01$ (dashed).}
    \label{fig:spectrum}
\end{figure}

The theoretical predictions of the aforementioned spectrum are depicted in Fig.~\ref{fig:spectrum}.
In this case, we set the model parameters to $A_\zeta=0.1$ (solid curves) and $A_\zeta=0.01$ (dashed curves).
The spectrum of the second order gravitational waves is represented in blue, while the combined spectrum of both the second and third order gravitational waves is shown in red.
In comparison to the second order gravitational waves, the third order gravitational waves primarily contribute to the spectrum around the peak frequencies.

{\emph{Joint constraints.}}
It is known that the \ac{PTA} probe is directly sensitive to the energy density of \acp{SIGW}. 
Following the methodology of Ref.~\cite{NANOGrav:2023hvm}, we examine the parameter space by conducting a Bayesian analysis over the \ac{NG15} dataset \cite{NANOGrav:2023gor}. 
The priors of $\log_{10} (f_\ast/\mathrm{Hz})$ and $\log A_{\zeta}$ are uniformly set within the intervals of [-11,-5] and [-3,1], respectively. 
In fact, our results are robust with respect to priors. 
Here, we neglect the very likely presence of an astrophysical foreground, which has been considered in Refs. \cite{Campeti:2020xwn,Poletti:2021ytu}.

\begin{figure}
\includegraphics[width =1 \columnwidth]{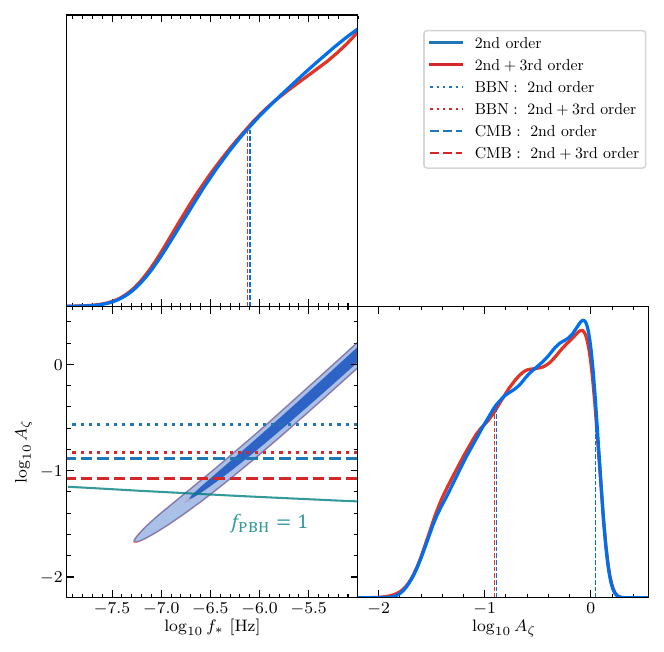}
\caption{The posteriors for the model parameters $A_\zeta$ and $f_\ast$ derived from the \ac{NG15} PTA dataset (contours), in addition to the \ac{BBN}  (thick dotted lines) and \ac{CMB} (thick dashed lines) upper limits on $A_\zeta$. For the one-dimensional posteriors, the 68\% confidence intervals are indicated by thin dashed lines. Deep green line denotes $f_{\mathrm{PBH}}=1$, indicating that all of dark matter is composed of \acp{PBH} \cite{NANOGrav:2023hvm}. }
\label{fig:posteriors}
\end{figure}

We consider two scenarios related to \acp{SIGW}. 
The first scenario (Scenario I) includes only the second order gravitational waves, while the second scenario (Scenario II) incorporates both the second  and third order gravitational waves. 
For both scenarios, we derive the posteriors of $f_{\ast}$ and $A_\zeta$, which are illustrated in Fig.~\ref{fig:posteriors}. 
Statistically, it is challenging to differentiate between the two scenarios since their posteriors nearly overlap.
We conclude that the third order gravitational waves, when being compared with the second order ones, have negligible impact on the interpretation of the observed \ac{PTA} signal in terms of \acp{SIGW}.

%
%
The \ac{BBN} and \ac{CMB} probes are indirectly sensitive to the energy density of \acp{SIGW} \footnote{Dr. Carlo Tasillo tells us via email that they have studied phase-transition gravitational waves by jointly analyzing the \ac{BBN}, \ac{CMB}, and NANOGrav 12.5-year data \cite{Bringmann:2023opz}.}. 
Specifically, they are sensitive only to the integrated energy-density fraction, as described by
\begin{equation}\label{eq:ogwh}
\int_{k_{\mathrm{min}}}^{\infty} d \ln k \ h^2 \Omega_{\mathrm{GW},0}(k) < 1.3\times10^{-6}\left(\frac{N_{\mathrm{eff}}-3.046}{0.234}\right) \, ,
\end{equation}
where $k_{\mathrm{min}}=2\pi f_{\mathrm{min}}$ sets the lower bound of the integral, and $N_{\mathrm{eff}}$ represents the number of relativistic species. As $f_{\mathrm{min}}$ is dependent on the physical process that occurred during the epochs of \ac{BBN} and \ac{CMB} formation, we adopt $f_{\mathrm{BBN}}\simeq1.5\times10^{-11}$ Hz for \ac{BBN} and $f_{\mathrm{CMB}}\simeq3\times10^{-17}$ Hz for \ac{CMB} \cite{Maggiore:2018sht}. According to the \texttt{Planck} 2018 \ac{CMB} plus BAO dataset \cite{Planck:2018vyg}, the right hand side of Eq.~(\ref{eq:ogwh}) equals $2.9\times10^{-7}$ \cite{Clarke:2020bil}, resulting in an upper limit of $A_\zeta\leq0.130$ for Scenario I and $A_\zeta\leq0.085$ for Scenario II. In contrast, for \ac{BBN} the right hand side equals $1.3\times10^{-6}$ \cite{Cooke:2013cba}, yielding an upper limit of $A_\zeta=0.275$ for Scenario I and $A_\zeta=0.150$ for Scenario II. These upper limits are illustrated in Fig.~\ref{fig:posteriors}. 
Though the contours are almost the same as those in Ref. \cite{NANOGrav:2023hvm}, the BBN and CMB upper bounds would significantly alter the posteriors via reducing a large portion of the posteriors. 
This indicates the importance of the data combination.

The results from the joint analysis are as follows. In both scenarios, the parameter region inferred from the \ac{NG15} data is notably refined by the inclusion of the \ac{BBN} and \ac{CMB} data. The permissible upper limit on $A_\zeta$ is somewhat smaller in Scenario II than in Scenario I, highlighting the significance of third order gravitational waves. 
Starting from the peak of the posterior, we derive the combined constraints on $A_\zeta$ and $f_\ast$ as 
\begin{eqnarray}
0.021<&A_\zeta&<0.085~, \label{eq:azeta}\label{eq:ares} \\ 5.0\times10^{-8}\mathrm{Hz}<&f_\ast&<5.0\times10^{-7}\mathrm{Hz}~,\label{eq:fres}\label{eq:fast}
\end{eqnarray}
at the 95\% confidence level. To our knowledge, these findings represent the state-of-the-art and most stringent constraints on the model parameters.

It should be noted that when $A_\zeta\gtrsim0.06$, the third order gravitational waves contribute more to the integrated energy density than the second order ones. 
This outcome suggests that the third order gravitational waves cannot be disregarded in the data analysis of \ac{BBN} and \ac{CMB}.

Taking into account both the second and third order gravitational waves, we also find that the \ac{CMB} bound, denoted by the dashed red line in Fig. \ref{fig:posteriors}, is comparable to the deep green line which indicates all of dark matter to be composed of \acp{PBH}, i.e., $f_{\mathrm{PBH}}=1$ \cite{NANOGrav:2023hvm}. 
There may be a risk of overproducing \acp{PBH}. 
However, the allowed maximum peak amplitude of the power spectrum is $A_{\zeta}\simeq0.058$, an amplitude making the third order contribution as nearly equal as the second order one. 
Therefore, it is important to take into account the third order gravitational waves in our analysis.

{\emph{Anticipated Constraints.}}
It is anticipated that the energy-density fraction spectrum of \acp{SIGW}, and subsequently the power spectrum of primordial curvature perturbations, will be potentially measured by the \ac{SKA} \cite{dewdney2009square,Weltman:2018zrl,Moore:2014lga}, $\mu$Ares \cite{Sesana:2019vho}, \acl{LISA} \cite{LISA:2017pwj,Robson:2018ifk}, \acl{BBO} \cite{Crowder:2005nr,Harry:2006fi}, \acl{DECIGO} \cite{Sato:2017dkf,Kawamura:2020pcg}, \acl{ET} \cite{Punturo:2010zz}, and Advanced LIGO and Virgo \cite{Harry:2010zz,VIRGO:2014yos,Somiya:2011np}.
Complementing \ac{CMB}, which is sensitive to the earliest stages of inflation, gravitational-wave probes offer the capability to investigate the physics of the early universe that occurred during later stages of inflation. Conducting multi-band gravitational-wave observations, we expect to explore the origin and evolution of the universe throughout the entire inflationary era.

Following Ref.~\cite{Zhao:2022kvz}, we examine this subject. During a \ac{GWB} search, if neglecting the very likely presence of an astrophysical foreground, the optimal \acl{SNR} is defined as \cite{Schmitz:2020syl}
\begin{equation}
\mathrm{SNR}^{2} = n_{\mathrm{det}}
T_{\mathrm{obs}} \left(\frac{3H_0^2}{2\pi^2}\right)^{2} \int_{f_{\mathrm{l}}}^{f_{\mathrm{u}}} \left[\frac{\Omega_{\mathrm{GW},0}(f)}{f^{3} S_{n}^{\mathrm{eff}}(f)}\right]^{2} {\rm d} f \ ,
\end{equation}
where $n_{\mathrm{det}}$ represents the number of detectors, $T_{\mathrm{obs}}$ is the observation duration, the frequency band extends from $f_{\mathrm{l}}$ to $f_{\mathrm{u}}$, and $S_{n}^{\mathrm{eff}}$ denotes the effective noise \acl{PSD} of the detector network. Here, $H_0=100h\ \mathrm{km}/\mathrm{s}/\mathrm{Mpc}$ is the Hubble constant. For the aforementioned experiments, we employ the setups summarized in Table 2 of Ref.~\cite{Campeti:2020xwn}.


\begin{figure}
\includegraphics[width =1 \columnwidth]{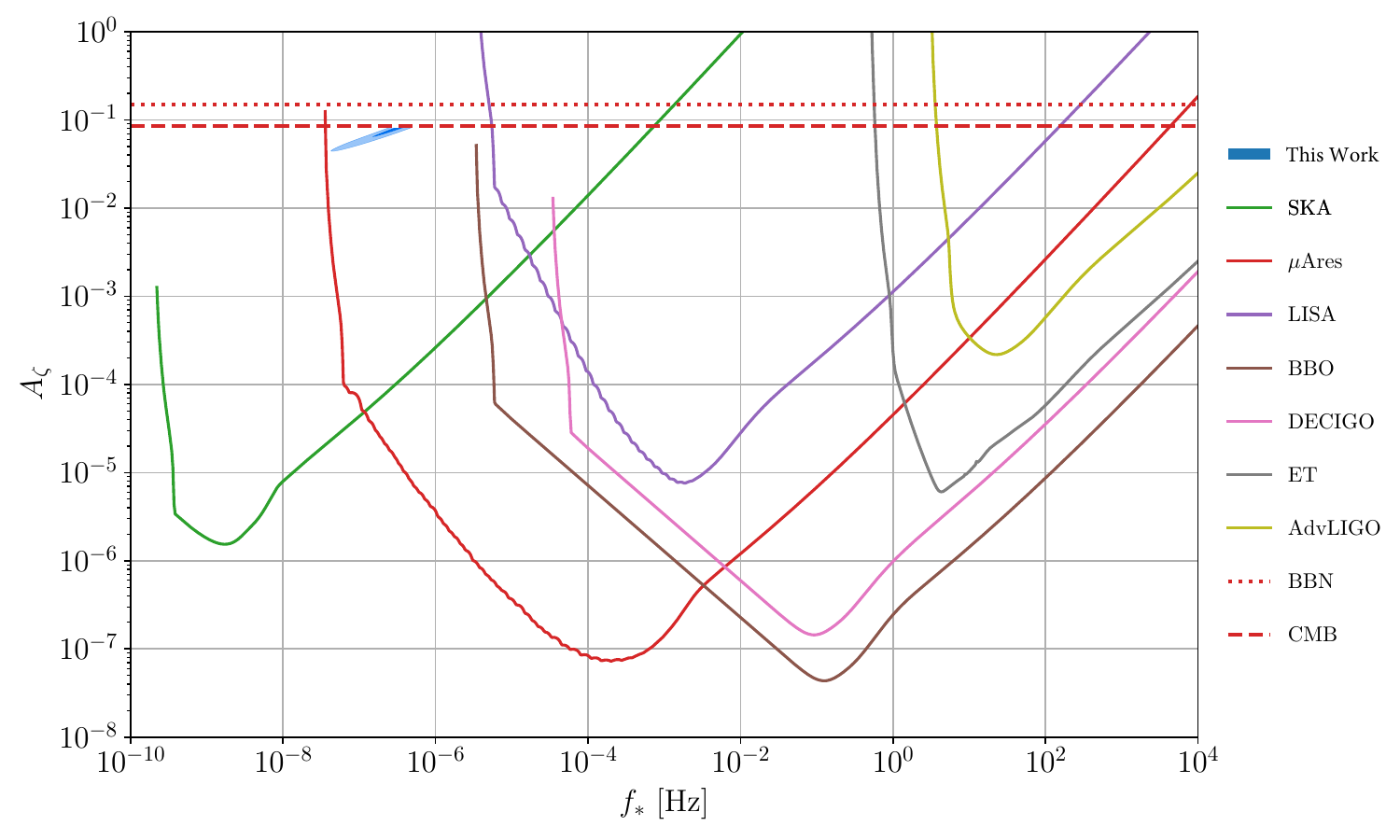}
\caption{Projected constraint capabilities of future gravitational-wave detection projects. The allowable parameter region inferred from Fig.~\ref{fig:posteriors}, as well as the upper limits from present \ac{BBN}  and \ac{CMB} observations, are also displayed for comparison. Here, we set $\mathrm{SNR}=1$ and neglect the very likely presence of an astrophysical foreground. }
\label{fig:contours}
\end{figure}

Requiring $\mathrm{SNR}=1$, we illustrate the anticipated constraint contours in the $A_\zeta$--$f_\ast$ plane for \ac{SKA} in Fig.~\ref{fig:contours}. The allowable parameter region (blue shaded contours) inferred from this study is presented for comparison. Notably, we find that our inferred contours can be further tested with \ac{SKA} and $\mu$Ares.
We note superb performance in measuring the primordial curvature spectral amplitude, i.e., $A_\zeta\sim10^{-6}$ for \ac{PTA} projects, $A_\zeta\sim10^{-8}$ for space-borne projects, and $A_\zeta\sim10^{-5}$ for ground-based projects. The high sensitivities should enable us to explore the early universe more comprehensively.

{\emph{Conclusion and Discussion.}}
In this study, we have delved into the gravitational waves induced by scalar perturbations, up to the third order, by scrutinizing recent \ac{PTA} datasets in conjunction with \ac{BBN}  and \ac{CMB} data. We have calculated the energy-density spectrum of \acp{SIGW} up to third order. Through the analysis of the joint datasets of \ac{BBN}, \ac{CMB}, and \ac{PTA}, we inferred the allowable parameter region, which was depicted in Fig.~\ref{fig:posteriors}. The inferred constraints on $A_\zeta$ and $f_\ast$ were presented in Eq.~(\ref{eq:ares}) and Eq.~(\ref{eq:fast}).

Interestingly, we found that the third order gravitational waves could contribute more to the integrated energy density than the second order ones when the primordial curvature spectral amplitude $A_\zeta$ exceeds approximately $0.06$. This finding underscores the importance of third order gravitational waves in our joint data analysis. 

As illustrated in Fig.~\ref{fig:contours}, 
we anticipated that the energy-density fraction spectrum of scalar-induced gravitational waves, and subsequently the power spectrum of primordial curvature perturbations, will be potentially measured by future gravitational-wave experiments, which should enable us to explore the early universe more comprehensively, and further test the predictions of our study. Our findings represent a significant step forward in our understanding of the universe, particularly in relation to cosmic inflation.

The next-generation \ac{CMB} experiments, e.g., CMB-S4 \cite{Abazajian:2019eic}, Simons Observatory \cite{SimonsObservatory:2018koc}, and LiteBIRD \cite{LiteBIRD:2022cnt}, are expected to reach better sensitivity that would lead to improvements of the present \ac{CMB} upper limits on $A_\zeta$, also indicating potential improvements of the best bound inferred in our current work.

Our results also suggest a mass range for \acp{PBH} (see reviews in Ref.~\cite{Carr:2020gox}) of $10^{-4.5}M_\odot<m_{\mathrm{PBH}}<10^{-2.5}M_\odot$, assuming the observed \ac{PTA} signal is interpreted as originating from \acp{SIGW} \cite{Kitajima:2023cek,Guo:2023hyp,Gouttenoire:2023ftk,Liu:2023ymk,Unal:2023srk,Unal:2023srk,Gouttenoire:2023nzr,Wang:2023ost,Huang:2023chx,Inomata:2023zup,Depta:2023qst,Franciolini:2023pbf,Han:2023olf,Choudhury:2023rks}. Notably, this mass range could account for the evidence for Planet 9 in the Outer Solar System \cite{2016AJ....151...22B,2019PhR...805....1B,2014Natur.507..471T,Scholtz:2019csj,2017Natur.548..183M,Niikura:2019kqi,Witten:2020ifl}. 


\begin{acknowledgements}
S.W. is partially supported by the National Natural Science Foundation of China (Grant No. 12175243), the National Key R\&D Program of China No. 2023YFC2206403, the Science Research Grants from the China Manned Space Project with No. CMS-CSST-2021-B01, and the Key Research Program of the Chinese Academy of Sciences (Grant No. XDPB15). Z.C.Z. is supported by the National Key Research and Development Program of China Grant No. 2021YFC2203001 and the National Natural Science Foundation of China (Grant NO. 12005016). Q.H.Z. is supported by the National Nature Science Foundation of China (Grant No.~12305073). This work is supported by High-performance Computing Platform of China Agricultural University.
\end{acknowledgements}

\begin{widetext}
\appendix

\section{Essential formulae \label{A}} 
 
Once the perturbed metric in Eq.~(\ref{1}) is known, we derive the equations of motion for \acp{SIGW} from the Einstein's equations in a hierarchical approach. 
For third order gravitational waves, we have \cite{Zhou:2021vcw}
\begin{equation}
h_{i j}^{(3)''}(\eta,\bm{x})+2 \mathcal{H}  h_{i j}^{(3)'}(\eta,\bm{x})-\Delta h_{i j}^{(3)}(\eta,\bm{x})=-12 \Lambda_{i j}^{l m} \mathcal{S}^{(3)}_{l m}(\eta,\bm{x})  \,, \label{4}
\end{equation}
where the source term is given by
\begin{equation}\label{5}
\begin{aligned}
			\Lambda^{ab}_{ij}S_{ab}^{(3)}(\eta,\bm{x}) = &  \Lambda^{ab}_{ij}\Bigg[ 12\phi^{(1)}\partial_a\phi^{(1)}\partial_b\phi^{(1)}-\frac{4}{\mathcal{H}}\phi^{(1)'}\partial_a\phi^{(1)}\partial_b\phi^{(1)}+\frac{2}{3\mathcal{H}^2}\Delta\phi^{(1)}\partial_a\phi^{(1)}\partial_b\phi^{(1)}\\
			&+\frac{2}{3\mathcal{H}^4}\Delta\phi^{(1)}\partial_a\phi^{(1)'}\partial_b\phi^{(1)'}-\frac{3}{\mathcal{H}^2}\phi^{(1)'}\partial_a\phi^{(1)'}\partial_b\phi^{(1)}-\frac{3}{\mathcal{H}^2}\phi^{(1)'}\partial_b\phi^{(1)'}\partial_a\phi^{(1)}\\
			&+\frac{2}{3\mathcal{H}^3}\Delta\phi^{(1)}\partial_a\phi^{(1)'}\partial_b\phi^{(1)}+\frac{2}{3\mathcal{H}^3}\Delta\phi^{(1)}\partial_b\phi^{(1)'}\partial_a\phi^{(1)}-\frac{2}{\mathcal{H}^3}\phi^{(1)'}\partial_a\phi^{(1)'}\partial_b\phi^{(1)'}-\frac{4}{\mathcal{H}^2}\phi^{(1)}\partial_a\phi^{(1)'}\partial_b\phi^{(1)'} \\
			&-\frac{1}{2}\phi^{(1)}\left( h_{ab}^{(2)''}+2 \mathcal{H}  h_{ab}^{(2)'}-\Delta h_{ab}^{(2)}\right)-\phi^{(1)}\Delta h_{ab}^{(2)}-\phi^{(1)'}\mathcal{H}h_{ab}^{(2)}-\frac{1}{3}\Delta \phi^{(1)}h_{ab}^{(2)}-\partial^c \phi^{(1)}\partial_c h_{ab}^{(2)} \\\
			&+\phi^{(1)}\partial_a\left(V_b^{(2)'}+2 \mathcal{H}V_b^{(2)} \right)+\phi^{(1)}\partial_b\left(V_a^{(2)'}+2 \mathcal{H}V_a^{(2)} \right)+\phi^{(1)'}\left(\partial_a V_b^{(2)}+\partial_b V_a^{(2)}\right)\\
			&-\frac{\phi^{(1)}}{8\mathcal{H}}\left(\partial_b\Delta V_a^{(2)}+\partial_a\Delta V_b^{(2)}\right)-\frac{\phi^{(1)'}}{8\mathcal{H}^2}\left(\partial_b\Delta V_a^{(2)}+\partial_a\Delta V_b^{(2)}\right) + \frac{1}{\mathcal{H}}\left(\phi^{(1)}\partial_a\partial_b\psi^{(2)'}\right) \\
			&+\frac{1}{\mathcal{H}}\left(\phi^{(1)'}\partial_a\partial_b\phi^{(2)}\right)+\frac{1}{\mathcal{H}^2}\left(\phi^{(1)'}\partial_a\partial_b\psi^{(2)'}\right)+3\left(\phi^{(1)}\partial_a\partial_b\phi^{(2)}\right)\Bigg] \, .
\end{aligned}
\end{equation}
Third order gravitational waves are produced by both the linear scalar perturbations and second order metric perturbations that are produced by the linear scalar perturbations. 
The latter were studied in the literature \cite{Inomata:2020cck,Zhou:2021vcw}.

In Fourier space, the evolution of the first and second order metric perturbations is given by 
\begin{subequations}\label{6-11}
\begin{eqnarray}
		\phi^{(1)}_{\bm k} &=&\Phi_{\rm k} T_\phi(k \eta)\label{7} \,, \\
		\phi^{(2)}_{\bm k} &=&  \int \frac{{\rm d}^3 k}{(2\pi)^3} \left\{ \Phi_{\bm k - p} \Phi_{\bm p}  I_\phi(|{\bm k}-{\bm p}|, |{\bm p}|,k\eta) \right\} \,, \label{8}\\
		\psi^{(2)}_{\bm k} &=&\int \frac{{\rm d}^3 k}{(2\pi)^3} \left\{ \Phi_{\bm k - \bm p} \Phi_{\bm p}  I_\psi(|{\bm k}-{\bm p}|, |{\bm p}|,k\eta) \right\} \,, \\
		V^{(2)}_{i,\bm k} &=&\int \frac{{\rm d}^3 k}{(2\pi)^3} \left\{ \Phi_{\bm k - \bm p} \Phi_{\bm p}  \mathcal{V}_i^{a b} \frac{p_a p_b}{k^2} I_V (|{\bm k}-{\bm p}|, |{\bm p}|,k\eta) \right\} \,, \\
		h^{(2)}_{ij,\bm k} &=&\int \frac{{\rm d}^3 k}{(2\pi)^3} \left\{ \Phi_{\bm k - p} \Phi_{\bm p} {\Lambda}_{ij}^{a b} \frac{p_a p_b}{k^2} I_h(|{\bm k}-{\bm p}|, |{\bm p}|,k\eta) \right\} \,, \label{11}
\end{eqnarray}
\end{subequations}
where $\mathcal{V}_j^{ab}(\bm k)=-i k_j(\delta^{ab}-k^ak^b/k^2)/k^2$ is helicity decomposition operator for vector perturbations, $\Lambda_{ij}^{ab}(\bm k)=\frac{1}{2}[(\delta^b_i-k^bk_i/k^2)(\delta^a_j-k^ak_j/k^2)+(\delta^a_i-k^ak_i/k^2)(\delta^b_j-k^bk_j/k^2)-(\delta^{ab}-k^ak^b/k^2)(\delta_{ij}-k_ik_j/k^2)]$ is that for tensor perturbations \cite{Chang:2020tji}, $\Phi_{\bm k}$ is a stochastic variable characterizing the primordial scalar perturbations. 
During radiation domination, the initial conditions lead to $\Phi_{\bm k}=2\zeta_{\bm k}/3$, where $\zeta_{\bm k}$ denotes primordial curvature perturbations with wavevector ${\bm k}$. 
The transfer function $T_\phi$ is obtained by solving the master equation for the linear scalar perturbations \cite{Dodelson:2003ft}. 
As shown in Ref.~\cite{Zhou:2021vcw}, the kernel functions $I_\phi$, $I_\psi$, $I_V$, and $I_h$ are obtained by solving the equations of motion for $\phi^{(2)}$, $\psi^{(2)}$, $V^{(2)}$, and $h^{(2)}$, respectively.

In Fourier space, Eq.~(\ref{5}) is reformulated as  
\begin{eqnarray}
\begin{aligned}
\Lambda^{ab}_{ij}(\bm k) S_{ab,\bm k}^{(3)}
= & \  \mathbb{P}_{(\phi^{(1)})^3,ij}(\bm k, \bm p, \bm q) f_1[\phi^{(1)}_{\bm k -\bm p},\phi^{(1)}_{\bm p - \bm q},\phi^{(1)}_{\bm q}]\\ &+\mathbb{P}_{\phi^{(1)}h^{(2)},ij}(\bm k, \bm p, \bm q) 
f_2[\phi^{(1)}_{\bm k -\bm p},\Phi_{\bm p - \bm q}\Phi_{\bm q} I_h(|\bm p -\bm q|, |\bm q|,|\bm p|\eta)] \\ & +\mathbb{P}_{\phi^{(1)}V^{(2)},ij}(\bm k, \bm p, \bm q) f_3[\phi^{(1)}_{\bm k -\bm p},\Phi_{\bm p - \bm q}\Phi_{\bm q} I_V(|\bm p -\bm q|, |\bm q|,|\bm p|\eta)]\\ & +\mathbb{P}_{\phi^{(1)}\psi^{(2)},ij}(\bm k, \bm p, \bm q) 
f_4[\phi^{(1)}_{\bm k -\bm p},\Phi_{\bm p - \bm q}\Phi_{\bm q} I_\psi(|\bm p -\bm q|, |\bm q|,|\bm p|\eta)]~, 
\end{aligned}
\end{eqnarray}    
where $f_n[...]$ with $n=1,2,3,4$ are homogeneous functions of order one, that can be read straightforwardly from Eq.~(\ref{5}), and we introduce 
\begin{subequations}\label{eqP}
\begin{eqnarray}
    &&\mathbb{P}_{(\phi^{(1)})^3,ij}(\bm k, \bm p, \bm q) = \frac{1}{k^2} \Lambda^{ab}_{ij}(\bm k)[(p_a-q_a)q_b+(p_b-q_b)q_a] \, ,\\
    &&\mathbb{P}_{\phi^{(1)}\psi^{(2)},ij}(\bm k, \bm p, \bm q) = \frac{1}{k^2} \Lambda^{ab}_{ij}(\bm k)p_a p_b 
 \, ,\\
    &&\mathbb{P}_{\phi^{(1)}V^{(2)},ij}(\bm k, \bm p, \bm q) = \frac{1}{k^2} \Lambda^{ab}_{ij}(\bm k)[\mathcal{V}^{cd}_a(\bm p)q_c q_d p_b + \mathcal{V}^{cd}_b(\bm p) q_c q_d p_a] \, ,\\
    &&\mathbb{P}_{\phi^{(1)}h^{(2)},ij}(\bm k, \bm p, \bm q) = \frac{1}{k^2} \Lambda^{ab}_{ij}(\bm k) \Lambda^{cd}_{ab}(\bm p) q_c q_d \, .
\end{eqnarray}
\end{subequations}

Solving Eq.~(\ref{4}) in Fourier space, we get the strain of third order gravitational waves as 
	\begin{eqnarray}
		\begin{aligned}
			h_{ij,\bm k}^{(3)} & \equiv  \int \frac{{\rm d}^3 p}{(2\pi)^3}\frac{{\rm d}^3 q}{(2\pi)^3}\Bigg\{ 
			\Phi_{\bm k - \bm p}\Phi_{\bm p - \bm q}\Phi_{\bm q}\, \Big[
			\mathbb{P}_{(\phi^{(1)})^3,ij} (\bm k, \bm p, \bm q)I^{(3)}_{(\phi^{(1)})^3} \left( | \bm{k} -  \bm{p} |, | \bm{p} - \bm{q} |, |  \bm{q} |, | \bm{p} |, k, \eta \right) \\
			& \qquad\qquad\qquad + \mathbb{P}_{\phi^{(1)}h^{(2)},ij} (\bm k, \bm p, \bm q)  I^{(3)}_{\phi^{(1)}h^{(2)}} \left( | \bm{k} -  \bm{p} |, | \bm{p} - \bm{q} |, |  \bm{q} |, | \bm{p} |, k, \eta \right) \\
			& \qquad\qquad\qquad  + \mathbb{P}_{\phi^{(1)}V^{(2)},ij} (\bm k, \bm p, \bm q)  I^{(3)}_{\phi^{(1)}V^{(2)}} \left( | \bm{k} -  \bm{p} |, | \bm{p} - \bm{q} |, |  \bm{q} |, | \bm{p} |, k, \eta \right) \\
			& \qquad\qquad\qquad +\mathbb{P}_{\phi^{(1)}\psi^{(2)},ij} (\bm k, \bm p, \bm q) I^{(3)}_{\phi^{(1)}\psi^{(2)}} \left( | \bm{k} -  \bm{p} |, | \bm{p} - \bm{q} |, |  \bm{q} |, | \bm{p} |, k, \eta \right)
			\Big] \Bigg\}~, 
		\end{aligned}\label{12}
	\end{eqnarray}     
where   
the kernel functions $I^{(3)}_{*}( | \bm{k} -  \bm{p} |, | \bm{p} - \bm{q} |, |  \bm{q} |, | \bm{p} |, k, \eta )$ are defined as 
\begin{subequations}
  \begin{eqnarray}
   && I^{(3)}_{(\phi^{(1)})^3} = \int ^\eta_0 {\rm d}\bar{\eta} G_k(\eta,\bar{\eta}) f_1[T_\phi(|\bm k - \bm p|\bar{\eta}),T_\phi(|\bm p - \bm q|\bar{\eta}),T_\phi(|\bm q|\bar{\eta})] ~,\\
   && I^{(3)}_{\phi^{(1)}h^{(2)}} =\int ^\eta_0 {\rm d}\bar{\eta} G_k(\eta,\bar{\eta}) f_2[T_\phi(|\bm k - \bm p|\bar{\eta}),I_h(|\bm p -\bm q|, |\bm q|,|\bm p|\bar{\eta})]  ~,\\
   && I^{(3)}_{\phi^{(1)}V^{(2)}} =\int ^\eta_0 {\rm d}\bar{\eta} G_k(\eta,\bar{\eta}) f_3[T_\phi(|\bm k - \bm p|\bar{\eta},I_V(|\bm p -\bm q|, |\bm q|,|\bm p|\bar{\eta})]  ~,\\
   && I^{(3)}_{\phi^{(1)}\psi^{(2)}} = \int ^\eta_0 {\rm d}\bar{\eta} G_k(\eta,\bar{\eta}) f_4[T_\phi(|\bm k - \bm p|\bar{\eta}),I_\psi(|\bm p -\bm q|, |\bm q|,|\bm p|\bar{\eta})] ~,
\end{eqnarray}  
\end{subequations}
with the Green's function during radiation domination being  
\begin{equation}
G_k(\eta,\bar{\eta})= \frac{1}{k}\sin[k(\eta-\bar{\eta})] \, .
\end{equation}
There is not a kernel function with a subscript $\phi^{(1)}\phi^{(2)}$, since $\phi^{(2)}$ has been replaced with $\psi^{(2)}$ via the second order equation of motion. 

\end{widetext}

\bibliography{PGW}

\begin{thebibliography}{84}%
\makeatletter
\providecommand \@ifxundefined [1]{%
 \@ifx{#1\undefined}
}%
\providecommand \@ifnum [1]{%
 \ifnum #1\expandafter \@firstoftwo
 \else \expandafter \@secondoftwo
 \fi
}%
\providecommand \@ifx [1]{%
 \ifx #1\expandafter \@firstoftwo
 \else \expandafter \@secondoftwo
 \fi
}%
\providecommand \natexlab [1]{#1}%
\providecommand \enquote  [1]{``#1''}%
\providecommand \bibnamefont  [1]{#1}%
\providecommand \bibfnamefont [1]{#1}%
\providecommand \citenamefont [1]{#1}%
\providecommand \href@noop [0]{\@secondoftwo}%
\providecommand \href [0]{\begingroup \@sanitize@url \@href}%
\providecommand \@href[1]{\@@startlink{#1}\@@href}%
\providecommand \@@href[1]{\endgroup#1\@@endlink}%
\providecommand \@sanitize@url [0]{\catcode `\\12\catcode `\$12\catcode
  `\&12\catcode `\#12\catcode `\^12\catcode `\_12\catcode `\%12\relax}%
\providecommand \@@startlink[1]{}%
\providecommand \@@endlink[0]{}%
\providecommand \url  [0]{\begingroup\@sanitize@url \@url }%
\providecommand \@url [1]{\endgroup\@href {#1}{\urlprefix }}%
\providecommand \urlprefix  [0]{URL }%
\providecommand \Eprint [0]{\href }%
\providecommand \doibase [0]{http://dx.doi.org/}%
\providecommand \selectlanguage [0]{\@gobble}%
\providecommand \bibinfo  [0]{\@secondoftwo}%
\providecommand \bibfield  [0]{\@secondoftwo}%
\providecommand \translation [1]{[#1]}%
\providecommand \BibitemOpen [0]{}%
\providecommand \bibitemStop [0]{}%
\providecommand \bibitemNoStop [0]{.\EOS\space}%
\providecommand \EOS [0]{\spacefactor3000\relax}%
\providecommand \BibitemShut  [1]{\csname bibitem#1\endcsname}%
\let\auto@bib@innerbib\@empty
\bibitem [{\citenamefont {Ananda}\ \emph {et~al.}(2007)\citenamefont {Ananda},
  \citenamefont {Clarkson},\ and\ \citenamefont {Wands}}]{Ananda:2006af}%
  \BibitemOpen
  \bibfield  {author} {\bibinfo {author} {\bibfnamefont {Kishore~N.}\
  \bibnamefont {Ananda}}, \bibinfo {author} {\bibfnamefont {Chris}\
  \bibnamefont {Clarkson}}, \ and\ \bibinfo {author} {\bibfnamefont {David}\
  \bibnamefont {Wands}},\ }\bibfield  {title} {\enquote {\bibinfo {title} {{The
  Cosmological gravitational wave background from primordial density
  perturbations}},}\ }\href {\doibase 10.1103/PhysRevD.75.123518} {\bibfield
  {journal} {\bibinfo  {journal} {Phys. Rev.}\ }\textbf {\bibinfo {volume}
  {D75}},\ \bibinfo {pages} {123518} (\bibinfo {year} {2007})},\ \Eprint
  {http://arxiv.org/abs/gr-qc/0612013} {arXiv:gr-qc/0612013 [gr-qc]}
  \BibitemShut {NoStop}%
\bibitem [{\citenamefont {Baumann}\ \emph {et~al.}(2007)\citenamefont
  {Baumann}, \citenamefont {Steinhardt}, \citenamefont {Takahashi},\ and\
  \citenamefont {Ichiki}}]{Baumann:2007zm}%
  \BibitemOpen
  \bibfield  {author} {\bibinfo {author} {\bibfnamefont {Daniel}\ \bibnamefont
  {Baumann}}, \bibinfo {author} {\bibfnamefont {Paul~J.}\ \bibnamefont
  {Steinhardt}}, \bibinfo {author} {\bibfnamefont {Keitaro}\ \bibnamefont
  {Takahashi}}, \ and\ \bibinfo {author} {\bibfnamefont {Kiyotomo}\
  \bibnamefont {Ichiki}},\ }\bibfield  {title} {\enquote {\bibinfo {title}
  {{Gravitational Wave Spectrum Induced by Primordial Scalar Perturbations}},}\
  }\href {\doibase 10.1103/PhysRevD.76.084019} {\bibfield  {journal} {\bibinfo
  {journal} {Phys. Rev.}\ }\textbf {\bibinfo {volume} {D76}},\ \bibinfo {pages}
  {084019} (\bibinfo {year} {2007})},\ \Eprint
  {http://arxiv.org/abs/hep-th/0703290} {arXiv:hep-th/0703290 [hep-th]}
  \BibitemShut {NoStop}%
\bibitem [{\citenamefont {Mollerach}\ \emph {et~al.}(2004)\citenamefont
  {Mollerach}, \citenamefont {Harari},\ and\ \citenamefont
  {Matarrese}}]{Mollerach:2003nq}%
  \BibitemOpen
  \bibfield  {author} {\bibinfo {author} {\bibfnamefont {Silvia}\ \bibnamefont
  {Mollerach}}, \bibinfo {author} {\bibfnamefont {Diego}\ \bibnamefont
  {Harari}}, \ and\ \bibinfo {author} {\bibfnamefont {Sabino}\ \bibnamefont
  {Matarrese}},\ }\bibfield  {title} {\enquote {\bibinfo {title} {{CMB
  polarization from secondary vector and tensor modes}},}\ }\href {\doibase
  10.1103/PhysRevD.69.063002} {\bibfield  {journal} {\bibinfo  {journal} {Phys.
  Rev.}\ }\textbf {\bibinfo {volume} {D69}},\ \bibinfo {pages} {063002}
  (\bibinfo {year} {2004})},\ \Eprint {http://arxiv.org/abs/astro-ph/0310711}
  {arXiv:astro-ph/0310711 [astro-ph]} \BibitemShut {NoStop}%
\bibitem [{\citenamefont {Assadullahi}\ and\ \citenamefont
  {Wands}(2010)}]{Assadullahi:2009jc}%
  \BibitemOpen
  \bibfield  {author} {\bibinfo {author} {\bibfnamefont {Hooshyar}\
  \bibnamefont {Assadullahi}}\ and\ \bibinfo {author} {\bibfnamefont {David}\
  \bibnamefont {Wands}},\ }\bibfield  {title} {\enquote {\bibinfo {title}
  {{Constraints on primordial density perturbations from induced gravitational
  waves}},}\ }\href {\doibase 10.1103/PhysRevD.81.023527} {\bibfield  {journal}
  {\bibinfo  {journal} {Phys. Rev.}\ }\textbf {\bibinfo {volume} {D81}},\
  \bibinfo {pages} {023527} (\bibinfo {year} {2010})},\ \Eprint
  {http://arxiv.org/abs/0907.4073} {arXiv:0907.4073 [astro-ph.CO]} \BibitemShut
  {NoStop}%
\bibitem [{\citenamefont {Espinosa}\ \emph {et~al.}(2018)\citenamefont
  {Espinosa}, \citenamefont {Racco},\ and\ \citenamefont
  {Riotto}}]{Espinosa:2018eve}%
  \BibitemOpen
  \bibfield  {author} {\bibinfo {author} {\bibfnamefont {Jos\'e~Ram\'on}\
  \bibnamefont {Espinosa}}, \bibinfo {author} {\bibfnamefont {Davide}\
  \bibnamefont {Racco}}, \ and\ \bibinfo {author} {\bibfnamefont {Antonio}\
  \bibnamefont {Riotto}},\ }\bibfield  {title} {\enquote {\bibinfo {title} {{A
  Cosmological Signature of the SM Higgs Instability: Gravitational Waves}},}\
  }\href {\doibase 10.1088/1475-7516/2018/09/012} {\bibfield  {journal}
  {\bibinfo  {journal} {JCAP}\ }\textbf {\bibinfo {volume} {09}},\ \bibinfo
  {pages} {012} (\bibinfo {year} {2018})},\ \Eprint
  {http://arxiv.org/abs/1804.07732} {arXiv:1804.07732 [hep-ph]} \BibitemShut
  {NoStop}%
\bibitem [{\citenamefont {Kohri}\ and\ \citenamefont
  {Terada}(2018)}]{Kohri:2018awv}%
  \BibitemOpen
  \bibfield  {author} {\bibinfo {author} {\bibfnamefont {Kazunori}\
  \bibnamefont {Kohri}}\ and\ \bibinfo {author} {\bibfnamefont {Takahiro}\
  \bibnamefont {Terada}},\ }\bibfield  {title} {\enquote {\bibinfo {title}
  {{Semianalytic calculation of gravitational wave spectrum nonlinearly induced
  from primordial curvature perturbations}},}\ }\href {\doibase
  10.1103/PhysRevD.97.123532} {\bibfield  {journal} {\bibinfo  {journal} {Phys.
  Rev.}\ }\textbf {\bibinfo {volume} {D97}},\ \bibinfo {pages} {123532}
  (\bibinfo {year} {2018})},\ \Eprint {http://arxiv.org/abs/1804.08577}
  {arXiv:1804.08577 [gr-qc]} \BibitemShut {NoStop}%
\bibitem [{\citenamefont {Xu}\ \emph {et~al.}(2023)\citenamefont {Xu} \emph
  {et~al.}}]{Xu:2023wog}%
  \BibitemOpen
  \bibfield  {author} {\bibinfo {author} {\bibfnamefont {Heng}\ \bibnamefont
  {Xu}} \emph {et~al.},\ }\bibfield  {title} {\enquote {\bibinfo {title}
  {{Searching for the Nano-Hertz Stochastic Gravitational Wave Background with
  the Chinese Pulsar Timing Array Data Release I}},}\ }\href {\doibase
  10.1088/1674-4527/acdfa5} {\bibfield  {journal} {\bibinfo  {journal} {Res.
  Astron. Astrophys.}\ }\textbf {\bibinfo {volume} {23}},\ \bibinfo {pages}
  {075024} (\bibinfo {year} {2023})},\ \Eprint
  {http://arxiv.org/abs/2306.16216} {arXiv:2306.16216 [astro-ph.HE]}
  \BibitemShut {NoStop}%
\bibitem [{\citenamefont {Antoniadis}\ \emph
  {et~al.}(2023{\natexlab{a}})\citenamefont {Antoniadis} \emph
  {et~al.}}]{Antoniadis:2023ott}%
  \BibitemOpen
  \bibfield  {author} {\bibinfo {author} {\bibfnamefont {J.}~\bibnamefont
  {Antoniadis}} \emph {et~al.},\ }\bibfield  {title} {\enquote {\bibinfo
  {title} {{The second data release from the European Pulsar Timing Array III.
  Search for gravitational wave signals}},}\ }\href@noop {} {\  (\bibinfo
  {year} {2023}{\natexlab{a}})},\ \Eprint {http://arxiv.org/abs/2306.16214}
  {arXiv:2306.16214 [astro-ph.HE]} \BibitemShut {NoStop}%
\bibitem [{\citenamefont {Agazie}\ \emph
  {et~al.}(2023{\natexlab{a}})\citenamefont {Agazie} \emph
  {et~al.}}]{NANOGrav:2023gor}%
  \BibitemOpen
  \bibfield  {author} {\bibinfo {author} {\bibfnamefont {Gabriella}\
  \bibnamefont {Agazie}} \emph {et~al.} (\bibinfo {collaboration} {NANOGrav}),\
  }\bibfield  {title} {\enquote {\bibinfo {title} {{The NANOGrav 15-year Data
  Set: Evidence for a Gravitational-Wave Background}},}\ }\href {\doibase
  10.3847/2041-8213/acdac6} {\bibfield  {journal} {\bibinfo  {journal}
  {Astrophys. J. Lett.}\ }\textbf {\bibinfo {volume} {951}} (\bibinfo {year}
  {2023}{\natexlab{a}}),\ 10.3847/2041-8213/acdac6},\ \Eprint
  {http://arxiv.org/abs/2306.16213} {arXiv:2306.16213 [astro-ph.HE]}
  \BibitemShut {NoStop}%
\bibitem [{\citenamefont {Reardon}\ \emph {et~al.}(2023)\citenamefont {Reardon}
  \emph {et~al.}}]{Reardon:2023gzh}%
  \BibitemOpen
  \bibfield  {author} {\bibinfo {author} {\bibfnamefont {Daniel~J.}\
  \bibnamefont {Reardon}} \emph {et~al.},\ }\bibfield  {title} {\enquote
  {\bibinfo {title} {{Search for an isotropic gravitational-wave background
  with the Parkes Pulsar Timing Array}},}\ }\href {\doibase
  10.3847/2041-8213/acdd02} {\bibfield  {journal} {\bibinfo  {journal}
  {Astrophys. J. Lett.}\ }\textbf {\bibinfo {volume} {951}} (\bibinfo {year}
  {2023}),\ 10.3847/2041-8213/acdd02},\ \Eprint
  {http://arxiv.org/abs/2306.16215} {arXiv:2306.16215 [astro-ph.HE]}
  \BibitemShut {NoStop}%
\bibitem [{\citenamefont {Antoniadis}\ \emph
  {et~al.}(2023{\natexlab{b}})\citenamefont {Antoniadis} \emph
  {et~al.}}]{Antoniadis:2023xlr}%
  \BibitemOpen
  \bibfield  {author} {\bibinfo {author} {\bibfnamefont {J.}~\bibnamefont
  {Antoniadis}} \emph {et~al.},\ }\bibfield  {title} {\enquote {\bibinfo
  {title} {{The second data release from the European Pulsar Timing Array: V.
  Implications for massive black holes, dark matter and the early Universe}},}\
  }\href@noop {} {\  (\bibinfo {year} {2023}{\natexlab{b}})},\ \Eprint
  {http://arxiv.org/abs/2306.16227} {arXiv:2306.16227 [astro-ph.CO]}
  \BibitemShut {NoStop}%
\bibitem [{\citenamefont {Afzal}\ \emph {et~al.}(2023)\citenamefont {Afzal}
  \emph {et~al.}}]{NANOGrav:2023hvm}%
  \BibitemOpen
  \bibfield  {author} {\bibinfo {author} {\bibfnamefont {Adeela}\ \bibnamefont
  {Afzal}} \emph {et~al.} (\bibinfo {collaboration} {NANOGrav}),\ }\bibfield
  {title} {\enquote {\bibinfo {title} {{The NANOGrav 15-year Data Set: Search
  for Signals from New Physics}},}\ }\href {\doibase 10.3847/2041-8213/acdc91}
  {\bibfield  {journal} {\bibinfo  {journal} {Astrophys. J. Lett.}\ }\textbf
  {\bibinfo {volume} {951}} (\bibinfo {year} {2023}),\
  10.3847/2041-8213/acdc91},\ \Eprint {http://arxiv.org/abs/2306.16219}
  {arXiv:2306.16219 [astro-ph.HE]} \BibitemShut {NoStop}%
\bibitem [{\citenamefont {Agazie}\ \emph
  {et~al.}(2023{\natexlab{b}})\citenamefont {Agazie} \emph
  {et~al.}}]{NANOGrav:2023hfp}%
  \BibitemOpen
  \bibfield  {author} {\bibinfo {author} {\bibfnamefont {Gabriella}\
  \bibnamefont {Agazie}} \emph {et~al.} (\bibinfo {collaboration} {NANOGrav}),\
  }\bibfield  {title} {\enquote {\bibinfo {title} {{The NANOGrav 15-year Data
  Set: Constraints on Supermassive Black Hole Binaries from the Gravitational
  Wave Background}},}\ }\href@noop {} {\  (\bibinfo {year}
  {2023}{\natexlab{b}})},\ \Eprint {http://arxiv.org/abs/2306.16220}
  {arXiv:2306.16220 [astro-ph.HE]} \BibitemShut {NoStop}%
\bibitem [{\citenamefont {Franciolini}\ \emph {et~al.}(2023)\citenamefont
  {Franciolini}, \citenamefont {Iovino}, \citenamefont {Vaskonen},\ and\
  \citenamefont {Veermae}}]{Franciolini:2023pbf}%
  \BibitemOpen
  \bibfield  {author} {\bibinfo {author} {\bibfnamefont {Gabriele}\
  \bibnamefont {Franciolini}}, \bibinfo {author} {\bibfnamefont {Antonio}\
  \bibnamefont {Iovino}, \bibfnamefont {Junior.}}, \bibinfo {author}
  {\bibfnamefont {Ville}\ \bibnamefont {Vaskonen}}, \ and\ \bibinfo {author}
  {\bibfnamefont {Hardi}\ \bibnamefont {Veermae}},\ }\bibfield  {title}
  {\enquote {\bibinfo {title} {{The recent gravitational wave observation by
  pulsar timing arrays and primordial black holes: the importance of
  non-gaussianities}},}\ }\href@noop {} {\  (\bibinfo {year} {2023})},\ \Eprint
  {http://arxiv.org/abs/2306.17149} {arXiv:2306.17149 [astro-ph.CO]}
  \BibitemShut {NoStop}%
\bibitem [{\citenamefont {Inomata}\ \emph {et~al.}(2023)\citenamefont
  {Inomata}, \citenamefont {Kohri},\ and\ \citenamefont
  {Terada}}]{Inomata:2023zup}%
  \BibitemOpen
  \bibfield  {author} {\bibinfo {author} {\bibfnamefont {Keisuke}\ \bibnamefont
  {Inomata}}, \bibinfo {author} {\bibfnamefont {Kazunori}\ \bibnamefont
  {Kohri}}, \ and\ \bibinfo {author} {\bibfnamefont {Takahiro}\ \bibnamefont
  {Terada}},\ }\bibfield  {title} {\enquote {\bibinfo {title} {{The Detected
  Stochastic Gravitational Waves and Sub-Solar Primordial Black Holes}},}\
  }\href@noop {} {\  (\bibinfo {year} {2023})},\ \Eprint
  {http://arxiv.org/abs/2306.17834} {arXiv:2306.17834 [astro-ph.CO]}
  \BibitemShut {NoStop}%
\bibitem [{\citenamefont {Wang}\ \emph {et~al.}(2023)\citenamefont {Wang},
  \citenamefont {Zhao}, \citenamefont {Li},\ and\ \citenamefont
  {Zhu}}]{Wang:2023ost}%
  \BibitemOpen
  \bibfield  {author} {\bibinfo {author} {\bibfnamefont {Sai}\ \bibnamefont
  {Wang}}, \bibinfo {author} {\bibfnamefont {Zhi-Chao}\ \bibnamefont {Zhao}},
  \bibinfo {author} {\bibfnamefont {Jun-Peng}\ \bibnamefont {Li}}, \ and\
  \bibinfo {author} {\bibfnamefont {Qing-Hua}\ \bibnamefont {Zhu}},\ }\bibfield
   {title} {\enquote {\bibinfo {title} {{Exploring the Implications of 2023
  Pulsar Timing Array Datasets for Scalar-Induced Gravitational Waves and
  Primordial Black Holes}},}\ }\href@noop {} {\  (\bibinfo {year} {2023})},\
  \Eprint {http://arxiv.org/abs/2307.00572} {arXiv:2307.00572 [astro-ph.CO]}
  \BibitemShut {NoStop}%
\bibitem [{\citenamefont {Liu}\ \emph {et~al.}(2023)\citenamefont {Liu},
  \citenamefont {Chen},\ and\ \citenamefont {Huang}}]{Liu:2023ymk}%
  \BibitemOpen
  \bibfield  {author} {\bibinfo {author} {\bibfnamefont {Lang}\ \bibnamefont
  {Liu}}, \bibinfo {author} {\bibfnamefont {Zu-Cheng}\ \bibnamefont {Chen}}, \
  and\ \bibinfo {author} {\bibfnamefont {Qing-Guo}\ \bibnamefont {Huang}},\
  }\bibfield  {title} {\enquote {\bibinfo {title} {{Implications for the
  non-Gaussianity of curvature perturbation from pulsar timing arrays}},}\
  }\href@noop {} {\  (\bibinfo {year} {2023})},\ \Eprint
  {http://arxiv.org/abs/2307.01102} {arXiv:2307.01102 [astro-ph.CO]}
  \BibitemShut {NoStop}%
\bibitem [{\citenamefont {Abe}\ and\ \citenamefont {Tada}(2023)}]{Abe:2023yrw}%
  \BibitemOpen
  \bibfield  {author} {\bibinfo {author} {\bibfnamefont {Katsuya~T.}\
  \bibnamefont {Abe}}\ and\ \bibinfo {author} {\bibfnamefont {Yuichiro}\
  \bibnamefont {Tada}},\ }\bibfield  {title} {\enquote {\bibinfo {title}
  {{Translating nano-Hertz gravitational wave background into primordial
  perturbations taking account of the cosmological QCD phase transition}},}\
  }\href@noop {} {\  (\bibinfo {year} {2023})},\ \Eprint
  {http://arxiv.org/abs/2307.01653} {arXiv:2307.01653 [astro-ph.CO]}
  \BibitemShut {NoStop}%
\bibitem [{\citenamefont {Ebadi}\ \emph {et~al.}(2023)\citenamefont {Ebadi},
  \citenamefont {Kumar}, \citenamefont {McCune}, \citenamefont {Tai},\ and\
  \citenamefont {Wang}}]{Ebadi:2023xhq}%
  \BibitemOpen
  \bibfield  {author} {\bibinfo {author} {\bibfnamefont {Reza}\ \bibnamefont
  {Ebadi}}, \bibinfo {author} {\bibfnamefont {Soubhik}\ \bibnamefont {Kumar}},
  \bibinfo {author} {\bibfnamefont {Amara}\ \bibnamefont {McCune}}, \bibinfo
  {author} {\bibfnamefont {Hanwen}\ \bibnamefont {Tai}}, \ and\ \bibinfo
  {author} {\bibfnamefont {Lian-Tao}\ \bibnamefont {Wang}},\ }\bibfield
  {title} {\enquote {\bibinfo {title} {{Gravitational Waves from Stochastic
  Scalar Fluctuations}},}\ }\href@noop {} {\  (\bibinfo {year} {2023})},\
  \Eprint {http://arxiv.org/abs/2307.01248} {arXiv:2307.01248 [astro-ph.CO]}
  \BibitemShut {NoStop}%
\bibitem [{\citenamefont {Figueroa}\ \emph {et~al.}(2023)\citenamefont
  {Figueroa}, \citenamefont {Pieroni}, \citenamefont {Ricciardone},\ and\
  \citenamefont {Simakachorn}}]{Figueroa:2023zhu}%
  \BibitemOpen
  \bibfield  {author} {\bibinfo {author} {\bibfnamefont {Daniel~G.}\
  \bibnamefont {Figueroa}}, \bibinfo {author} {\bibfnamefont {Mauro}\
  \bibnamefont {Pieroni}}, \bibinfo {author} {\bibfnamefont {Angelo}\
  \bibnamefont {Ricciardone}}, \ and\ \bibinfo {author} {\bibfnamefont {Peera}\
  \bibnamefont {Simakachorn}},\ }\bibfield  {title} {\enquote {\bibinfo {title}
  {{Cosmological Background Interpretation of Pulsar Timing Array Data}},}\
  }\href@noop {} {\  (\bibinfo {year} {2023})},\ \Eprint
  {http://arxiv.org/abs/2307.02399} {arXiv:2307.02399 [astro-ph.CO]}
  \BibitemShut {NoStop}%
\bibitem [{\citenamefont {Yi}\ \emph {et~al.}(2023)\citenamefont {Yi},
  \citenamefont {Gao}, \citenamefont {Gong}, \citenamefont {Wang},\ and\
  \citenamefont {Zhang}}]{Yi:2023mbm}%
  \BibitemOpen
  \bibfield  {author} {\bibinfo {author} {\bibfnamefont {Zhu}\ \bibnamefont
  {Yi}}, \bibinfo {author} {\bibfnamefont {Qing}\ \bibnamefont {Gao}}, \bibinfo
  {author} {\bibfnamefont {Yungui}\ \bibnamefont {Gong}}, \bibinfo {author}
  {\bibfnamefont {Yue}\ \bibnamefont {Wang}}, \ and\ \bibinfo {author}
  {\bibfnamefont {Fengge}\ \bibnamefont {Zhang}},\ }\bibfield  {title}
  {\enquote {\bibinfo {title} {{The waveform of the scalar induced
  gravitational waves in light of Pulsar Timing Array data}},}\ }\href@noop {}
  {\  (\bibinfo {year} {2023})},\ \Eprint {http://arxiv.org/abs/2307.02467}
  {arXiv:2307.02467 [gr-qc]} \BibitemShut {NoStop}%
\bibitem [{\citenamefont {Madge}\ \emph {et~al.}(2023)\citenamefont {Madge},
  \citenamefont {Morgante}, \citenamefont {Ib\'a\~nez}, \citenamefont
  {Ramberg},\ and\ \citenamefont {Schenk}}]{Madge:2023cak}%
  \BibitemOpen
  \bibfield  {author} {\bibinfo {author} {\bibfnamefont {Eric}\ \bibnamefont
  {Madge}}, \bibinfo {author} {\bibfnamefont {Enrico}\ \bibnamefont
  {Morgante}}, \bibinfo {author} {\bibfnamefont {Cristina~Puchades}\
  \bibnamefont {Ib\'a\~nez}}, \bibinfo {author} {\bibfnamefont {Nicklas}\
  \bibnamefont {Ramberg}}, \ and\ \bibinfo {author} {\bibfnamefont {Sebastian}\
  \bibnamefont {Schenk}},\ }\bibfield  {title} {\enquote {\bibinfo {title}
  {{Primordial gravitational waves in the nano-Hertz regime and PTA data --
  towards solving the GW inverse problem}},}\ }\href@noop {} {\  (\bibinfo
  {year} {2023})},\ \Eprint {http://arxiv.org/abs/2306.14856} {arXiv:2306.14856
  [hep-ph]} \BibitemShut {NoStop}%
\bibitem [{\citenamefont {Cai}\ \emph {et~al.}(2023)\citenamefont {Cai},
  \citenamefont {He}, \citenamefont {Ma}, \citenamefont {Yan},\ and\
  \citenamefont {Yuan}}]{Cai:2023dls}%
  \BibitemOpen
  \bibfield  {author} {\bibinfo {author} {\bibfnamefont {Yi-Fu}\ \bibnamefont
  {Cai}}, \bibinfo {author} {\bibfnamefont {Xin-Chen}\ \bibnamefont {He}},
  \bibinfo {author} {\bibfnamefont {Xiaohan}\ \bibnamefont {Ma}}, \bibinfo
  {author} {\bibfnamefont {Sheng-Feng}\ \bibnamefont {Yan}}, \ and\ \bibinfo
  {author} {\bibfnamefont {Guan-Wen}\ \bibnamefont {Yuan}},\ }\bibfield
  {title} {\enquote {\bibinfo {title} {{Limits on scalar-induced gravitational
  waves from the stochastic background by pulsar timing array observations}},}\
  }\href@noop {} {\  (\bibinfo {year} {2023})},\ \Eprint
  {http://arxiv.org/abs/2306.17822} {arXiv:2306.17822 [gr-qc]} \BibitemShut
  {NoStop}%
\bibitem [{\citenamefont {Maggiore}(2018)}]{Maggiore:2018sht}%
  \BibitemOpen
  \bibfield  {author} {\bibinfo {author} {\bibfnamefont {Michele}\ \bibnamefont
  {Maggiore}},\ }\href@noop {} {\emph {\bibinfo {title} {{Gravitational Waves.
  Vol. 2: Astrophysics and Cosmology}}}}\ (\bibinfo  {publisher} {Oxford
  University Press},\ \bibinfo {year} {2018})\BibitemShut {NoStop}%
\bibitem [{\citenamefont {Moore}\ and\ \citenamefont
  {Vecchio}(2021)}]{Moore:2021ibq}%
  \BibitemOpen
  \bibfield  {author} {\bibinfo {author} {\bibfnamefont {Christopher~J.}\
  \bibnamefont {Moore}}\ and\ \bibinfo {author} {\bibfnamefont {Alberto}\
  \bibnamefont {Vecchio}},\ }\bibfield  {title} {\enquote {\bibinfo {title}
  {{Ultra-low-frequency gravitational waves from cosmological and astrophysical
  processes}},}\ }\href {\doibase 10.1038/s41550-021-01489-8} {\bibfield
  {journal} {\bibinfo  {journal} {Nature Astron.}\ }\textbf {\bibinfo {volume}
  {5}},\ \bibinfo {pages} {1268--1274} (\bibinfo {year} {2021})},\ \Eprint
  {http://arxiv.org/abs/2104.15130} {arXiv:2104.15130 [astro-ph.CO]}
  \BibitemShut {NoStop}%
\bibitem [{\citenamefont {Smith}\ \emph {et~al.}(2006)\citenamefont {Smith},
  \citenamefont {Pierpaoli},\ and\ \citenamefont
  {Kamionkowski}}]{Smith:2006nka}%
  \BibitemOpen
  \bibfield  {author} {\bibinfo {author} {\bibfnamefont {Tristan~L.}\
  \bibnamefont {Smith}}, \bibinfo {author} {\bibfnamefont {Elena}\ \bibnamefont
  {Pierpaoli}}, \ and\ \bibinfo {author} {\bibfnamefont {Marc}\ \bibnamefont
  {Kamionkowski}},\ }\bibfield  {title} {\enquote {\bibinfo {title} {{A new
  cosmic microwave background constraint to primordial gravitational waves}},}\
  }\href {\doibase 10.1103/PhysRevLett.97.021301} {\bibfield  {journal}
  {\bibinfo  {journal} {Phys. Rev. Lett.}\ }\textbf {\bibinfo {volume} {97}},\
  \bibinfo {pages} {021301} (\bibinfo {year} {2006})},\ \Eprint
  {http://arxiv.org/abs/astro-ph/0603144} {arXiv:astro-ph/0603144} \BibitemShut
  {NoStop}%
\bibitem [{\citenamefont {Clarke}\ \emph {et~al.}(2020)\citenamefont {Clarke},
  \citenamefont {Copeland},\ and\ \citenamefont {Moss}}]{Clarke:2020bil}%
  \BibitemOpen
  \bibfield  {author} {\bibinfo {author} {\bibfnamefont {Thomas~J.}\
  \bibnamefont {Clarke}}, \bibinfo {author} {\bibfnamefont {Edmund~J.}\
  \bibnamefont {Copeland}}, \ and\ \bibinfo {author} {\bibfnamefont {Adam}\
  \bibnamefont {Moss}},\ }\bibfield  {title} {\enquote {\bibinfo {title}
  {{Constraints on primordial gravitational waves from the Cosmic Microwave
  Background}},}\ }\href {\doibase 10.1088/1475-7516/2020/10/002} {\bibfield
  {journal} {\bibinfo  {journal} {JCAP}\ }\textbf {\bibinfo {volume} {10}},\
  \bibinfo {pages} {002} (\bibinfo {year} {2020})},\ \Eprint
  {http://arxiv.org/abs/2004.11396} {arXiv:2004.11396 [astro-ph.CO]}
  \BibitemShut {NoStop}%
\bibitem [{\citenamefont {Cooke}\ \emph {et~al.}(2014)\citenamefont {Cooke},
  \citenamefont {Pettini}, \citenamefont {Jorgenson}, \citenamefont {Murphy},\
  and\ \citenamefont {Steidel}}]{Cooke:2013cba}%
  \BibitemOpen
  \bibfield  {author} {\bibinfo {author} {\bibfnamefont {Ryan}\ \bibnamefont
  {Cooke}}, \bibinfo {author} {\bibfnamefont {Max}\ \bibnamefont {Pettini}},
  \bibinfo {author} {\bibfnamefont {Regina~A.}\ \bibnamefont {Jorgenson}},
  \bibinfo {author} {\bibfnamefont {Michael~T.}\ \bibnamefont {Murphy}}, \ and\
  \bibinfo {author} {\bibfnamefont {Charles~C.}\ \bibnamefont {Steidel}},\
  }\bibfield  {title} {\enquote {\bibinfo {title} {{Precision measures of the
  primordial abundance of deuterium}},}\ }\href {\doibase
  10.1088/0004-637X/781/1/31} {\bibfield  {journal} {\bibinfo  {journal}
  {Astrophys. J.}\ }\textbf {\bibinfo {volume} {781}},\ \bibinfo {pages} {31}
  (\bibinfo {year} {2014})},\ \Eprint {http://arxiv.org/abs/1308.3240}
  {arXiv:1308.3240 [astro-ph.CO]} \BibitemShut {NoStop}%
\bibitem [{\citenamefont {Yuan}\ \emph {et~al.}(2019)\citenamefont {Yuan},
  \citenamefont {Chen},\ and\ \citenamefont {Huang}}]{Yuan:2019udt}%
  \BibitemOpen
  \bibfield  {author} {\bibinfo {author} {\bibfnamefont {Chen}\ \bibnamefont
  {Yuan}}, \bibinfo {author} {\bibfnamefont {Zu-Cheng}\ \bibnamefont {Chen}}, \
  and\ \bibinfo {author} {\bibfnamefont {Qing-Guo}\ \bibnamefont {Huang}},\
  }\bibfield  {title} {\enquote {\bibinfo {title} {{Probing
  primordial\textendash{}black-hole dark matter with scalar induced
  gravitational waves}},}\ }\href {\doibase 10.1103/PhysRevD.100.081301}
  {\bibfield  {journal} {\bibinfo  {journal} {Phys. Rev. D}\ }\textbf {\bibinfo
  {volume} {100}},\ \bibinfo {pages} {081301} (\bibinfo {year} {2019})},\
  \Eprint {http://arxiv.org/abs/1906.11549} {arXiv:1906.11549 [astro-ph.CO]}
  \BibitemShut {NoStop}%
\bibitem [{\citenamefont {Zhou}\ \emph {et~al.}(2022)\citenamefont {Zhou},
  \citenamefont {Zhang}, \citenamefont {Zhu},\ and\ \citenamefont
  {Chang}}]{Zhou:2021vcw}%
  \BibitemOpen
  \bibfield  {author} {\bibinfo {author} {\bibfnamefont {Jing-Zhi}\
  \bibnamefont {Zhou}}, \bibinfo {author} {\bibfnamefont {Xukun}\ \bibnamefont
  {Zhang}}, \bibinfo {author} {\bibfnamefont {Qing-Hua}\ \bibnamefont {Zhu}}, \
  and\ \bibinfo {author} {\bibfnamefont {Zhe}\ \bibnamefont {Chang}},\
  }\bibfield  {title} {\enquote {\bibinfo {title} {{The third order scalar
  induced gravitational waves}},}\ }\href {\doibase
  10.1088/1475-7516/2022/05/013} {\bibfield  {journal} {\bibinfo  {journal}
  {JCAP}\ }\textbf {\bibinfo {volume} {05}},\ \bibinfo {pages} {013} (\bibinfo
  {year} {2022})},\ \Eprint {http://arxiv.org/abs/2106.01641} {arXiv:2106.01641
  [astro-ph.CO]} \BibitemShut {NoStop}%
\bibitem [{\citenamefont {Chang}\ \emph {et~al.}(2021)\citenamefont {Chang},
  \citenamefont {Wang},\ and\ \citenamefont {Zhu}}]{Chang:2020tji}%
  \BibitemOpen
  \bibfield  {author} {\bibinfo {author} {\bibfnamefont {Zhe}\ \bibnamefont
  {Chang}}, \bibinfo {author} {\bibfnamefont {Sai}\ \bibnamefont {Wang}}, \
  and\ \bibinfo {author} {\bibfnamefont {Qing-Hua}\ \bibnamefont {Zhu}},\
  }\bibfield  {title} {\enquote {\bibinfo {title} {{Note on gauge invariance of
  second order cosmological perturbations}},}\ }\href {\doibase
  10.1088/1674-1137/ac0c74} {\bibfield  {journal} {\bibinfo  {journal} {Chin.
  Phys. C}\ }\textbf {\bibinfo {volume} {45}},\ \bibinfo {pages} {095101}
  (\bibinfo {year} {2021})},\ \Eprint {http://arxiv.org/abs/2009.11025}
  {arXiv:2009.11025 [astro-ph.CO]} \BibitemShut {NoStop}%
\bibitem [{\citenamefont {Ivanov}\ \emph {et~al.}(1994)\citenamefont {Ivanov},
  \citenamefont {Naselsky},\ and\ \citenamefont {Novikov}}]{Ivanov:1994pa}%
  \BibitemOpen
  \bibfield  {author} {\bibinfo {author} {\bibfnamefont {P.}~\bibnamefont
  {Ivanov}}, \bibinfo {author} {\bibfnamefont {P.}~\bibnamefont {Naselsky}}, \
  and\ \bibinfo {author} {\bibfnamefont {I.}~\bibnamefont {Novikov}},\
  }\bibfield  {title} {\enquote {\bibinfo {title} {{Inflation and primordial
  black holes as dark matter}},}\ }\href {\doibase 10.1103/PhysRevD.50.7173}
  {\bibfield  {journal} {\bibinfo  {journal} {Phys. Rev. D}\ }\textbf {\bibinfo
  {volume} {50}},\ \bibinfo {pages} {7173--7178} (\bibinfo {year}
  {1994})}\BibitemShut {NoStop}%
\bibitem [{\citenamefont {Garcia-Bellido}\ \emph {et~al.}(1996)\citenamefont
  {Garcia-Bellido}, \citenamefont {Linde},\ and\ \citenamefont
  {Wands}}]{Garcia-Bellido:1996mdl}%
  \BibitemOpen
  \bibfield  {author} {\bibinfo {author} {\bibfnamefont {Juan}\ \bibnamefont
  {Garcia-Bellido}}, \bibinfo {author} {\bibfnamefont {Andrei~D.}\ \bibnamefont
  {Linde}}, \ and\ \bibinfo {author} {\bibfnamefont {David}\ \bibnamefont
  {Wands}},\ }\bibfield  {title} {\enquote {\bibinfo {title} {{Density
  perturbations and black hole formation in hybrid inflation}},}\ }\href
  {\doibase 10.1103/PhysRevD.54.6040} {\bibfield  {journal} {\bibinfo
  {journal} {Phys. Rev. D}\ }\textbf {\bibinfo {volume} {54}},\ \bibinfo
  {pages} {6040--6058} (\bibinfo {year} {1996})},\ \Eprint
  {http://arxiv.org/abs/astro-ph/9605094} {arXiv:astro-ph/9605094} \BibitemShut
  {NoStop}%
\bibitem [{\citenamefont {Yokoyama}(1998)}]{Yokoyama:1998pt}%
  \BibitemOpen
  \bibfield  {author} {\bibinfo {author} {\bibfnamefont {Jun'ichi}\
  \bibnamefont {Yokoyama}},\ }\bibfield  {title} {\enquote {\bibinfo {title}
  {{Chaotic new inflation and formation of primordial black holes}},}\ }\href
  {\doibase 10.1103/PhysRevD.58.083510} {\bibfield  {journal} {\bibinfo
  {journal} {Phys. Rev. D}\ }\textbf {\bibinfo {volume} {58}},\ \bibinfo
  {pages} {083510} (\bibinfo {year} {1998})},\ \Eprint
  {http://arxiv.org/abs/astro-ph/9802357} {arXiv:astro-ph/9802357} \BibitemShut
  {NoStop}%
\bibitem [{\citenamefont {Cai}\ \emph {et~al.}(2018)\citenamefont {Cai},
  \citenamefont {Tong}, \citenamefont {Wang},\ and\ \citenamefont
  {Yan}}]{Cai:2018tuh}%
  \BibitemOpen
  \bibfield  {author} {\bibinfo {author} {\bibfnamefont {Yi-Fu}\ \bibnamefont
  {Cai}}, \bibinfo {author} {\bibfnamefont {Xi}~\bibnamefont {Tong}}, \bibinfo
  {author} {\bibfnamefont {Dong-Gang}\ \bibnamefont {Wang}}, \ and\ \bibinfo
  {author} {\bibfnamefont {Sheng-Feng}\ \bibnamefont {Yan}},\ }\bibfield
  {title} {\enquote {\bibinfo {title} {{Primordial Black Holes from Sound Speed
  Resonance during Inflation}},}\ }\href {\doibase
  10.1103/PhysRevLett.121.081306} {\bibfield  {journal} {\bibinfo  {journal}
  {Phys. Rev. Lett.}\ }\textbf {\bibinfo {volume} {121}},\ \bibinfo {pages}
  {081306} (\bibinfo {year} {2018})},\ \Eprint
  {http://arxiv.org/abs/1805.03639} {arXiv:1805.03639 [astro-ph.CO]}
  \BibitemShut {NoStop}%
\bibitem [{\citenamefont {Cai}\ \emph {et~al.}(2019)\citenamefont {Cai},
  \citenamefont {Chen}, \citenamefont {Tong}, \citenamefont {Wang},\ and\
  \citenamefont {Yan}}]{Cai:2019jah}%
  \BibitemOpen
  \bibfield  {author} {\bibinfo {author} {\bibfnamefont {Yi-Fu}\ \bibnamefont
  {Cai}}, \bibinfo {author} {\bibfnamefont {Chao}\ \bibnamefont {Chen}},
  \bibinfo {author} {\bibfnamefont {Xi}~\bibnamefont {Tong}}, \bibinfo {author}
  {\bibfnamefont {Dong-Gang}\ \bibnamefont {Wang}}, \ and\ \bibinfo {author}
  {\bibfnamefont {Sheng-Feng}\ \bibnamefont {Yan}},\ }\bibfield  {title}
  {\enquote {\bibinfo {title} {{When Primordial Black Holes from Sound Speed
  Resonance Meet a Stochastic Background of Gravitational Waves}},}\ }\href
  {\doibase 10.1103/PhysRevD.100.043518} {\bibfield  {journal} {\bibinfo
  {journal} {Phys. Rev. D}\ }\textbf {\bibinfo {volume} {100}},\ \bibinfo
  {pages} {043518} (\bibinfo {year} {2019})},\ \Eprint
  {http://arxiv.org/abs/1902.08187} {arXiv:1902.08187 [astro-ph.CO]}
  \BibitemShut {NoStop}%
\bibitem [{\citenamefont {Cai}\ \emph {et~al.}(2020)\citenamefont {Cai},
  \citenamefont {Guo}, \citenamefont {Liu}, \citenamefont {Liu},\ and\
  \citenamefont {Yang}}]{Cai:2019bmk}%
  \BibitemOpen
  \bibfield  {author} {\bibinfo {author} {\bibfnamefont {Rong-Gen}\
  \bibnamefont {Cai}}, \bibinfo {author} {\bibfnamefont {Zong-Kuan}\
  \bibnamefont {Guo}}, \bibinfo {author} {\bibfnamefont {Jing}\ \bibnamefont
  {Liu}}, \bibinfo {author} {\bibfnamefont {Lang}\ \bibnamefont {Liu}}, \ and\
  \bibinfo {author} {\bibfnamefont {Xing-Yu}\ \bibnamefont {Yang}},\ }\bibfield
   {title} {\enquote {\bibinfo {title} {{Primordial black holes and
  gravitational waves from parametric amplification of curvature
  perturbations}},}\ }\href {\doibase 10.1088/1475-7516/2020/06/013} {\bibfield
   {journal} {\bibinfo  {journal} {JCAP}\ }\textbf {\bibinfo {volume} {06}},\
  \bibinfo {pages} {013} (\bibinfo {year} {2020})},\ \Eprint
  {http://arxiv.org/abs/1912.10437} {arXiv:1912.10437 [astro-ph.CO]}
  \BibitemShut {NoStop}%
\bibitem [{\citenamefont {Chen}\ and\ \citenamefont
  {Cai}(2019)}]{Chen:2019zza}%
  \BibitemOpen
  \bibfield  {author} {\bibinfo {author} {\bibfnamefont {Chao}\ \bibnamefont
  {Chen}}\ and\ \bibinfo {author} {\bibfnamefont {Yi-Fu}\ \bibnamefont {Cai}},\
  }\bibfield  {title} {\enquote {\bibinfo {title} {{Primordial black holes from
  sound speed resonance in the inflaton-curvaton mixed scenario}},}\ }\href
  {\doibase 10.1088/1475-7516/2019/10/068} {\bibfield  {journal} {\bibinfo
  {journal} {JCAP}\ }\textbf {\bibinfo {volume} {10}},\ \bibinfo {pages} {068}
  (\bibinfo {year} {2019})},\ \Eprint {http://arxiv.org/abs/1908.03942}
  {arXiv:1908.03942 [astro-ph.CO]} \BibitemShut {NoStop}%
\bibitem [{\citenamefont {Chen}\ \emph {et~al.}(2020)\citenamefont {Chen},
  \citenamefont {Ma},\ and\ \citenamefont {Cai}}]{Chen:2020uhe}%
  \BibitemOpen
  \bibfield  {author} {\bibinfo {author} {\bibfnamefont {Chao}\ \bibnamefont
  {Chen}}, \bibinfo {author} {\bibfnamefont {Xiao-Han}\ \bibnamefont {Ma}}, \
  and\ \bibinfo {author} {\bibfnamefont {Yi-Fu}\ \bibnamefont {Cai}},\
  }\bibfield  {title} {\enquote {\bibinfo {title} {{Dirac-Born-Infeld
  realization of sound speed resonance mechanism for primordial black
  holes}},}\ }\href {\doibase 10.1103/PhysRevD.102.063526} {\bibfield
  {journal} {\bibinfo  {journal} {Phys. Rev. D}\ }\textbf {\bibinfo {volume}
  {102}},\ \bibinfo {pages} {063526} (\bibinfo {year} {2020})},\ \Eprint
  {http://arxiv.org/abs/2003.03821} {arXiv:2003.03821 [astro-ph.CO]}
  \BibitemShut {NoStop}%
\bibitem [{\citenamefont {Yu}\ and\ \citenamefont {Wang}(2023)}]{Yu:2023jrs}%
  \BibitemOpen
  \bibfield  {author} {\bibinfo {author} {\bibfnamefont {Yan-Heng}\
  \bibnamefont {Yu}}\ and\ \bibinfo {author} {\bibfnamefont {Sai}\ \bibnamefont
  {Wang}},\ }\bibfield  {title} {\enquote {\bibinfo {title} {{Anisotropies in
  Scalar-Induced Gravitational-Wave Background from Inflaton-Curvaton Mixed
  Scenario with Sound Speed Resonance}},}\ }\href@noop {} {\  (\bibinfo {year}
  {2023})},\ \Eprint {http://arxiv.org/abs/2310.14606} {arXiv:2310.14606
  [astro-ph.CO]} \BibitemShut {NoStop}%
\bibitem [{\citenamefont {Wang}\ \emph {et~al.}(2019)\citenamefont {Wang},
  \citenamefont {Terada},\ and\ \citenamefont {Kohri}}]{Wang:2019kaf}%
  \BibitemOpen
  \bibfield  {author} {\bibinfo {author} {\bibfnamefont {Sai}\ \bibnamefont
  {Wang}}, \bibinfo {author} {\bibfnamefont {Takahiro}\ \bibnamefont {Terada}},
  \ and\ \bibinfo {author} {\bibfnamefont {Kazunori}\ \bibnamefont {Kohri}},\
  }\bibfield  {title} {\enquote {\bibinfo {title} {{Prospective constraints on
  the primordial black hole abundance from the stochastic gravitational-wave
  backgrounds produced by coalescing events and curvature perturbations}},}\
  }\href {\doibase 10.1103/PhysRevD.99.103531} {\bibfield  {journal} {\bibinfo
  {journal} {Phys. Rev. D}\ }\textbf {\bibinfo {volume} {99}},\ \bibinfo
  {pages} {103531} (\bibinfo {year} {2019})},\ \bibinfo {note} {[Erratum:
  Phys.Rev.D 101, 069901 (2020)]},\ \Eprint {http://arxiv.org/abs/1903.05924}
  {arXiv:1903.05924 [astro-ph.CO]} \BibitemShut {NoStop}%
\bibitem [{\citenamefont {Aghanim}\ \emph {et~al.}(2020)\citenamefont {Aghanim}
  \emph {et~al.}}]{Planck:2018vyg}%
  \BibitemOpen
  \bibfield  {author} {\bibinfo {author} {\bibfnamefont {N.}~\bibnamefont
  {Aghanim}} \emph {et~al.} (\bibinfo {collaboration} {Planck}),\ }\bibfield
  {title} {\enquote {\bibinfo {title} {{Planck 2018 results. VI. Cosmological
  parameters}},}\ }\href {\doibase 10.1051/0004-6361/201833910} {\bibfield
  {journal} {\bibinfo  {journal} {Astron. Astrophys.}\ }\textbf {\bibinfo
  {volume} {641}},\ \bibinfo {pages} {A6} (\bibinfo {year} {2020})},\ \bibinfo
  {note} {[Erratum: Astron.Astrophys. 652, C4 (2021)]},\ \Eprint
  {http://arxiv.org/abs/1807.06209} {arXiv:1807.06209 [astro-ph.CO]}
  \BibitemShut {NoStop}%
\bibitem [{\citenamefont {Saikawa}\ and\ \citenamefont
  {Shirai}(2018)}]{Saikawa:2018rcs}%
  \BibitemOpen
  \bibfield  {author} {\bibinfo {author} {\bibfnamefont {Ken'ichi}\
  \bibnamefont {Saikawa}}\ and\ \bibinfo {author} {\bibfnamefont {Satoshi}\
  \bibnamefont {Shirai}},\ }\bibfield  {title} {\enquote {\bibinfo {title}
  {{Primordial gravitational waves, precisely: The role of thermodynamics in
  the Standard Model}},}\ }\href {\doibase 10.1088/1475-7516/2018/05/035}
  {\bibfield  {journal} {\bibinfo  {journal} {JCAP}\ }\textbf {\bibinfo
  {volume} {1805}},\ \bibinfo {pages} {035} (\bibinfo {year} {2018})},\ \Eprint
  {http://arxiv.org/abs/1803.01038} {arXiv:1803.01038 [hep-ph]} \BibitemShut
  {NoStop}%
\bibitem [{\citenamefont {Campeti}\ \emph {et~al.}(2021)\citenamefont
  {Campeti}, \citenamefont {Komatsu}, \citenamefont {Poletti},\ and\
  \citenamefont {Baccigalupi}}]{Campeti:2020xwn}%
  \BibitemOpen
  \bibfield  {author} {\bibinfo {author} {\bibfnamefont {Paolo}\ \bibnamefont
  {Campeti}}, \bibinfo {author} {\bibfnamefont {Eiichiro}\ \bibnamefont
  {Komatsu}}, \bibinfo {author} {\bibfnamefont {Davide}\ \bibnamefont
  {Poletti}}, \ and\ \bibinfo {author} {\bibfnamefont {Carlo}\ \bibnamefont
  {Baccigalupi}},\ }\bibfield  {title} {\enquote {\bibinfo {title} {{Measuring
  the spectrum of primordial gravitational waves with CMB, PTA and Laser
  Interferometers}},}\ }\href {\doibase 10.1088/1475-7516/2021/01/012}
  {\bibfield  {journal} {\bibinfo  {journal} {JCAP}\ }\textbf {\bibinfo
  {volume} {01}},\ \bibinfo {pages} {012} (\bibinfo {year} {2021})},\ \Eprint
  {http://arxiv.org/abs/2007.04241} {arXiv:2007.04241 [astro-ph.CO]}
  \BibitemShut {NoStop}%
\bibitem [{\citenamefont {Poletti}(2021)}]{Poletti:2021ytu}%
  \BibitemOpen
  \bibfield  {author} {\bibinfo {author} {\bibfnamefont {Davide}\ \bibnamefont
  {Poletti}},\ }\bibfield  {title} {\enquote {\bibinfo {title} {{Measuring the
  primordial gravitational wave background in the presence of other stochastic
  signals}},}\ }\href {\doibase 10.1088/1475-7516/2021/05/052} {\bibfield
  {journal} {\bibinfo  {journal} {JCAP}\ }\textbf {\bibinfo {volume} {05}},\
  \bibinfo {pages} {052} (\bibinfo {year} {2021})},\ \Eprint
  {http://arxiv.org/abs/2101.02713} {arXiv:2101.02713 [gr-qc]} \BibitemShut
  {NoStop}%
\bibitem [{\citenamefont {Bringmann}\ \emph {et~al.}(2023)\citenamefont
  {Bringmann}, \citenamefont {Depta}, \citenamefont {Konstandin}, \citenamefont
  {Schmidt-Hoberg},\ and\ \citenamefont {Tasillo}}]{Bringmann:2023opz}%
  \BibitemOpen
  \bibfield  {author} {\bibinfo {author} {\bibfnamefont {Torsten}\ \bibnamefont
  {Bringmann}}, \bibinfo {author} {\bibfnamefont {Paul~Frederik}\ \bibnamefont
  {Depta}}, \bibinfo {author} {\bibfnamefont {Thomas}\ \bibnamefont
  {Konstandin}}, \bibinfo {author} {\bibfnamefont {Kai}\ \bibnamefont
  {Schmidt-Hoberg}}, \ and\ \bibinfo {author} {\bibfnamefont {Carlo}\
  \bibnamefont {Tasillo}},\ }\bibfield  {title} {\enquote {\bibinfo {title}
  {{Does NANOGrav observe a dark sector phase transition?}}}\ }\href@noop {} {\
   (\bibinfo {year} {2023})},\ \Eprint {http://arxiv.org/abs/2306.09411}
  {arXiv:2306.09411 [astro-ph.CO]} \BibitemShut {NoStop}%
\bibitem [{\citenamefont {Dewdney}\ \emph {et~al.}(2009)\citenamefont
  {Dewdney}, \citenamefont {Hall}, \citenamefont {Schilizzi},\ and\
  \citenamefont {Lazio}}]{dewdney2009square}%
  \BibitemOpen
  \bibfield  {author} {\bibinfo {author} {\bibfnamefont {Peter~E}\ \bibnamefont
  {Dewdney}}, \bibinfo {author} {\bibfnamefont {Peter~J}\ \bibnamefont {Hall}},
  \bibinfo {author} {\bibfnamefont {Richard~T}\ \bibnamefont {Schilizzi}}, \
  and\ \bibinfo {author} {\bibfnamefont {T~Joseph~LW}\ \bibnamefont {Lazio}},\
  }\bibfield  {title} {\enquote {\bibinfo {title} {The square kilometre
  array},}\ }\href@noop {} {\bibfield  {journal} {\bibinfo  {journal}
  {Proceedings of the IEEE}\ }\textbf {\bibinfo {volume} {97}},\ \bibinfo
  {pages} {1482--1496} (\bibinfo {year} {2009})}\BibitemShut {NoStop}%
\bibitem [{\citenamefont {Weltman}\ \emph {et~al.}(2020)\citenamefont {Weltman}
  \emph {et~al.}}]{Weltman:2018zrl}%
  \BibitemOpen
  \bibfield  {author} {\bibinfo {author} {\bibfnamefont {A.}~\bibnamefont
  {Weltman}} \emph {et~al.},\ }\bibfield  {title} {\enquote {\bibinfo {title}
  {{Fundamental physics with the Square Kilometre Array}},}\ }\href {\doibase
  10.1017/pasa.2019.42} {\bibfield  {journal} {\bibinfo  {journal} {Publ.
  Astron. Soc. Austral.}\ }\textbf {\bibinfo {volume} {37}},\ \bibinfo {pages}
  {e002} (\bibinfo {year} {2020})},\ \Eprint {http://arxiv.org/abs/1810.02680}
  {arXiv:1810.02680 [astro-ph.CO]} \BibitemShut {NoStop}%
\bibitem [{\citenamefont {Moore}\ \emph {et~al.}(2015)\citenamefont {Moore},
  \citenamefont {Cole},\ and\ \citenamefont {Berry}}]{Moore:2014lga}%
  \BibitemOpen
  \bibfield  {author} {\bibinfo {author} {\bibfnamefont {C.~J.}\ \bibnamefont
  {Moore}}, \bibinfo {author} {\bibfnamefont {R.~H.}\ \bibnamefont {Cole}}, \
  and\ \bibinfo {author} {\bibfnamefont {C.~P.~L.}\ \bibnamefont {Berry}},\
  }\bibfield  {title} {\enquote {\bibinfo {title} {{Gravitational-wave
  sensitivity curves}},}\ }\href {\doibase 10.1088/0264-9381/32/1/015014}
  {\bibfield  {journal} {\bibinfo  {journal} {Class. Quant. Grav.}\ }\textbf
  {\bibinfo {volume} {32}},\ \bibinfo {pages} {015014} (\bibinfo {year}
  {2015})},\ \Eprint {http://arxiv.org/abs/1408.0740} {arXiv:1408.0740 [gr-qc]}
  \BibitemShut {NoStop}%
\bibitem [{\citenamefont {Sesana}\ \emph {et~al.}(2021)\citenamefont {Sesana}
  \emph {et~al.}}]{Sesana:2019vho}%
  \BibitemOpen
  \bibfield  {author} {\bibinfo {author} {\bibfnamefont {Alberto}\ \bibnamefont
  {Sesana}} \emph {et~al.},\ }\bibfield  {title} {\enquote {\bibinfo {title}
  {{Unveiling the gravitational universe at $\mu$-Hz frequencies}},}\ }\href
  {\doibase 10.1007/s10686-021-09709-9} {\bibfield  {journal} {\bibinfo
  {journal} {Exper. Astron.}\ }\textbf {\bibinfo {volume} {51}},\ \bibinfo
  {pages} {1333--1383} (\bibinfo {year} {2021})},\ \Eprint
  {http://arxiv.org/abs/1908.11391} {arXiv:1908.11391 [astro-ph.IM]}
  \BibitemShut {NoStop}%
\bibitem [{\citenamefont {Amaro-Seoane}\ \emph {et~al.}(2017)\citenamefont
  {Amaro-Seoane} \emph {et~al.}}]{LISA:2017pwj}%
  \BibitemOpen
  \bibfield  {author} {\bibinfo {author} {\bibfnamefont {Pau}\ \bibnamefont
  {Amaro-Seoane}} \emph {et~al.} (\bibinfo {collaboration} {LISA}),\ }\bibfield
   {title} {\enquote {\bibinfo {title} {{Laser Interferometer Space
  Antenna}},}\ }\href@noop {} {\  (\bibinfo {year} {2017})},\ \Eprint
  {http://arxiv.org/abs/1702.00786} {arXiv:1702.00786 [astro-ph.IM]}
  \BibitemShut {NoStop}%
\bibitem [{\citenamefont {Robson}\ \emph {et~al.}(2019)\citenamefont {Robson},
  \citenamefont {Cornish},\ and\ \citenamefont {Liu}}]{Robson:2018ifk}%
  \BibitemOpen
  \bibfield  {author} {\bibinfo {author} {\bibfnamefont {Travis}\ \bibnamefont
  {Robson}}, \bibinfo {author} {\bibfnamefont {Neil~J.}\ \bibnamefont
  {Cornish}}, \ and\ \bibinfo {author} {\bibfnamefont {Chang}\ \bibnamefont
  {Liu}},\ }\bibfield  {title} {\enquote {\bibinfo {title} {{The construction
  and use of LISA sensitivity curves}},}\ }\href {\doibase
  10.1088/1361-6382/ab1101} {\bibfield  {journal} {\bibinfo  {journal} {Class.
  Quant. Grav.}\ }\textbf {\bibinfo {volume} {36}},\ \bibinfo {pages} {105011}
  (\bibinfo {year} {2019})},\ \Eprint {http://arxiv.org/abs/1803.01944}
  {arXiv:1803.01944 [astro-ph.HE]} \BibitemShut {NoStop}%
\bibitem [{\citenamefont {Crowder}\ and\ \citenamefont
  {Cornish}(2005)}]{Crowder:2005nr}%
  \BibitemOpen
  \bibfield  {author} {\bibinfo {author} {\bibfnamefont {Jeff}\ \bibnamefont
  {Crowder}}\ and\ \bibinfo {author} {\bibfnamefont {Neil~J.}\ \bibnamefont
  {Cornish}},\ }\bibfield  {title} {\enquote {\bibinfo {title} {{Beyond LISA:
  Exploring future gravitational wave missions}},}\ }\href {\doibase
  10.1103/PhysRevD.72.083005} {\bibfield  {journal} {\bibinfo  {journal} {Phys.
  Rev. D}\ }\textbf {\bibinfo {volume} {72}},\ \bibinfo {pages} {083005}
  (\bibinfo {year} {2005})},\ \Eprint {http://arxiv.org/abs/gr-qc/0506015}
  {arXiv:gr-qc/0506015} \BibitemShut {NoStop}%
\bibitem [{\citenamefont {Harry}\ \emph {et~al.}(2006)\citenamefont {Harry},
  \citenamefont {Fritschel}, \citenamefont {Shaddock}, \citenamefont
  {Folkner},\ and\ \citenamefont {Phinney}}]{Harry:2006fi}%
  \BibitemOpen
  \bibfield  {author} {\bibinfo {author} {\bibfnamefont {G.~M.}\ \bibnamefont
  {Harry}}, \bibinfo {author} {\bibfnamefont {P.}~\bibnamefont {Fritschel}},
  \bibinfo {author} {\bibfnamefont {D.~A.}\ \bibnamefont {Shaddock}}, \bibinfo
  {author} {\bibfnamefont {W.}~\bibnamefont {Folkner}}, \ and\ \bibinfo
  {author} {\bibfnamefont {E.~S.}\ \bibnamefont {Phinney}},\ }\bibfield
  {title} {\enquote {\bibinfo {title} {{Laser interferometry for the big bang
  observer}},}\ }\href {\doibase 10.1088/0264-9381/23/15/008} {\bibfield
  {journal} {\bibinfo  {journal} {Class. Quant. Grav.}\ }\textbf {\bibinfo
  {volume} {23}},\ \bibinfo {pages} {4887--4894} (\bibinfo {year} {2006})},\
  \bibinfo {note} {[Erratum: Class.Quant.Grav. 23, 7361 (2006)]}\BibitemShut
  {NoStop}%
\bibitem [{\citenamefont {Sato}\ \emph {et~al.}(2017)\citenamefont {Sato} \emph
  {et~al.}}]{Sato:2017dkf}%
  \BibitemOpen
  \bibfield  {author} {\bibinfo {author} {\bibfnamefont {Shuichi}\ \bibnamefont
  {Sato}} \emph {et~al.},\ }\bibfield  {title} {\enquote {\bibinfo {title}
  {{The status of DECIGO}},}\ }\bibfield  {booktitle} {\emph {\bibinfo
  {booktitle} {{Proceedings, 11th International LISA Symposium: Zurich,
  Switzerland, September 5-9, 2016}}},\ }\href {\doibase
  10.1088/1742-6596/840/1/012010} {\bibfield  {journal} {\bibinfo  {journal}
  {J. Phys. Conf. Ser.}\ }\textbf {\bibinfo {volume} {840}},\ \bibinfo {pages}
  {012010} (\bibinfo {year} {2017})}\BibitemShut {NoStop}%
\bibitem [{\citenamefont {Kawamura}\ \emph {et~al.}(2021)\citenamefont
  {Kawamura} \emph {et~al.}}]{Kawamura:2020pcg}%
  \BibitemOpen
  \bibfield  {author} {\bibinfo {author} {\bibfnamefont {Seiji}\ \bibnamefont
  {Kawamura}} \emph {et~al.},\ }\bibfield  {title} {\enquote {\bibinfo {title}
  {{Current status of space gravitational wave antenna DECIGO and B-DECIGO}},}\
  }\href {\doibase 10.1093/ptep/ptab019} {\bibfield  {journal} {\bibinfo
  {journal} {PTEP}\ }\textbf {\bibinfo {volume} {2021}},\ \bibinfo {pages}
  {05A105} (\bibinfo {year} {2021})},\ \Eprint
  {http://arxiv.org/abs/2006.13545} {arXiv:2006.13545 [gr-qc]} \BibitemShut
  {NoStop}%
\bibitem [{\citenamefont {Punturo}\ \emph {et~al.}(2010)\citenamefont {Punturo}
  \emph {et~al.}}]{Punturo:2010zz}%
  \BibitemOpen
  \bibfield  {author} {\bibinfo {author} {\bibfnamefont {M.}~\bibnamefont
  {Punturo}} \emph {et~al.},\ }\bibfield  {title} {\enquote {\bibinfo {title}
  {{The Einstein Telescope: A third-generation gravitational wave
  observatory}},}\ }\bibfield  {booktitle} {\emph {\bibinfo {booktitle}
  {{Proceedings, 14th Workshop on Gravitational wave data analysis (GWDAW-14):
  Rome, Italy, January 26-29, 2010}}},\ }\href {\doibase
  10.1088/0264-9381/27/19/194002} {\bibfield  {journal} {\bibinfo  {journal}
  {Class. Quant. Grav.}\ }\textbf {\bibinfo {volume} {27}},\ \bibinfo {pages}
  {194002} (\bibinfo {year} {2010})}\BibitemShut {NoStop}%
\bibitem [{\citenamefont {Harry}(2010)}]{Harry:2010zz}%
  \BibitemOpen
  \bibfield  {author} {\bibinfo {author} {\bibfnamefont {Gregory~M.}\
  \bibnamefont {Harry}} (\bibinfo {collaboration} {LIGO Scientific}),\
  }\bibfield  {title} {\enquote {\bibinfo {title} {{Advanced LIGO: The next
  generation of gravitational wave detectors}},}\ }\href {\doibase
  10.1088/0264-9381/27/8/084006} {\bibfield  {journal} {\bibinfo  {journal}
  {Class. Quant. Grav.}\ }\textbf {\bibinfo {volume} {27}},\ \bibinfo {pages}
  {084006} (\bibinfo {year} {2010})}\BibitemShut {NoStop}%
\bibitem [{\citenamefont {Acernese}\ \emph {et~al.}(2015)\citenamefont
  {Acernese} \emph {et~al.}}]{VIRGO:2014yos}%
  \BibitemOpen
  \bibfield  {author} {\bibinfo {author} {\bibfnamefont {F.}~\bibnamefont
  {Acernese}} \emph {et~al.} (\bibinfo {collaboration} {VIRGO}),\ }\bibfield
  {title} {\enquote {\bibinfo {title} {{Advanced Virgo: a second-generation
  interferometric gravitational wave detector}},}\ }\href {\doibase
  10.1088/0264-9381/32/2/024001} {\bibfield  {journal} {\bibinfo  {journal}
  {Class. Quant. Grav.}\ }\textbf {\bibinfo {volume} {32}},\ \bibinfo {pages}
  {024001} (\bibinfo {year} {2015})},\ \Eprint {http://arxiv.org/abs/1408.3978}
  {arXiv:1408.3978 [gr-qc]} \BibitemShut {NoStop}%
\bibitem [{\citenamefont {Somiya}(2012)}]{Somiya:2011np}%
  \BibitemOpen
  \bibfield  {author} {\bibinfo {author} {\bibfnamefont {Kentaro}\ \bibnamefont
  {Somiya}} (\bibinfo {collaboration} {KAGRA}),\ }\bibfield  {title} {\enquote
  {\bibinfo {title} {{Detector configuration of KAGRA: The Japanese cryogenic
  gravitational-wave detector}},}\ }\href {\doibase
  10.1088/0264-9381/29/12/124007} {\bibfield  {journal} {\bibinfo  {journal}
  {Class. Quant. Grav.}\ }\textbf {\bibinfo {volume} {29}},\ \bibinfo {pages}
  {124007} (\bibinfo {year} {2012})},\ \Eprint {http://arxiv.org/abs/1111.7185}
  {arXiv:1111.7185 [gr-qc]} \BibitemShut {NoStop}%
\bibitem [{\citenamefont {Zhao}\ and\ \citenamefont
  {Wang}(2023)}]{Zhao:2022kvz}%
  \BibitemOpen
  \bibfield  {author} {\bibinfo {author} {\bibfnamefont {Zhi-Chao}\
  \bibnamefont {Zhao}}\ and\ \bibinfo {author} {\bibfnamefont {Sai}\
  \bibnamefont {Wang}},\ }\bibfield  {title} {\enquote {\bibinfo {title}
  {{Bayesian Implications for the Primordial Black Holes from
  NANOGrav\textquoteright{}s Pulsar-Timing Data Using the Scalar-Induced
  Gravitational Waves}},}\ }\href {\doibase 10.3390/universe9040157} {\bibfield
   {journal} {\bibinfo  {journal} {Universe}\ }\textbf {\bibinfo {volume}
  {9}},\ \bibinfo {pages} {157} (\bibinfo {year} {2023})},\ \Eprint
  {http://arxiv.org/abs/2211.09450} {arXiv:2211.09450 [astro-ph.CO]}
  \BibitemShut {NoStop}%
\bibitem [{\citenamefont {Schmitz}(2021)}]{Schmitz:2020syl}%
  \BibitemOpen
  \bibfield  {author} {\bibinfo {author} {\bibfnamefont {Kai}\ \bibnamefont
  {Schmitz}},\ }\bibfield  {title} {\enquote {\bibinfo {title} {{New
  Sensitivity Curves for Gravitational-Wave Signals from Cosmological Phase
  Transitions}},}\ }\href {\doibase 10.1007/JHEP01(2021)097} {\bibfield
  {journal} {\bibinfo  {journal} {JHEP}\ }\textbf {\bibinfo {volume} {01}},\
  \bibinfo {pages} {097} (\bibinfo {year} {2021})},\ \Eprint
  {http://arxiv.org/abs/2002.04615} {arXiv:2002.04615 [hep-ph]} \BibitemShut
  {NoStop}%
\bibitem [{\citenamefont {Abazajian}\ \emph {et~al.}(2019)\citenamefont
  {Abazajian} \emph {et~al.}}]{Abazajian:2019eic}%
  \BibitemOpen
  \bibfield  {author} {\bibinfo {author} {\bibfnamefont {Kevork}\ \bibnamefont
  {Abazajian}} \emph {et~al.},\ }\bibfield  {title} {\enquote {\bibinfo {title}
  {{CMB-S4 Science Case, Reference Design, and Project Plan}},}\ }\href@noop {}
  {\  (\bibinfo {year} {2019})},\ \Eprint {http://arxiv.org/abs/1907.04473}
  {arXiv:1907.04473 [astro-ph.IM]} \BibitemShut {NoStop}%
\bibitem [{\citenamefont {Ade}\ \emph {et~al.}(2019)\citenamefont {Ade} \emph
  {et~al.}}]{SimonsObservatory:2018koc}%
  \BibitemOpen
  \bibfield  {author} {\bibinfo {author} {\bibfnamefont {Peter}\ \bibnamefont
  {Ade}} \emph {et~al.} (\bibinfo {collaboration} {Simons Observatory}),\
  }\bibfield  {title} {\enquote {\bibinfo {title} {{The Simons Observatory:
  Science goals and forecasts}},}\ }\href {\doibase
  10.1088/1475-7516/2019/02/056} {\bibfield  {journal} {\bibinfo  {journal}
  {JCAP}\ }\textbf {\bibinfo {volume} {02}},\ \bibinfo {pages} {056} (\bibinfo
  {year} {2019})},\ \Eprint {http://arxiv.org/abs/1808.07445} {arXiv:1808.07445
  [astro-ph.CO]} \BibitemShut {NoStop}%
\bibitem [{\citenamefont {Allys}\ \emph {et~al.}(2023)\citenamefont {Allys}
  \emph {et~al.}}]{LiteBIRD:2022cnt}%
  \BibitemOpen
  \bibfield  {author} {\bibinfo {author} {\bibfnamefont {E.}~\bibnamefont
  {Allys}} \emph {et~al.} (\bibinfo {collaboration} {LiteBIRD}),\ }\bibfield
  {title} {\enquote {\bibinfo {title} {{Probing Cosmic Inflation with the
  LiteBIRD Cosmic Microwave Background Polarization Survey}},}\ }\href
  {\doibase 10.1093/ptep/ptac150} {\bibfield  {journal} {\bibinfo  {journal}
  {PTEP}\ }\textbf {\bibinfo {volume} {2023}},\ \bibinfo {pages} {042F01}
  (\bibinfo {year} {2023})},\ \Eprint {http://arxiv.org/abs/2202.02773}
  {arXiv:2202.02773 [astro-ph.IM]} \BibitemShut {NoStop}%
\bibitem [{\citenamefont {Carr}\ \emph {et~al.}(2021)\citenamefont {Carr},
  \citenamefont {Kohri}, \citenamefont {Sendouda},\ and\ \citenamefont
  {Yokoyama}}]{Carr:2020gox}%
  \BibitemOpen
  \bibfield  {author} {\bibinfo {author} {\bibfnamefont {Bernard}\ \bibnamefont
  {Carr}}, \bibinfo {author} {\bibfnamefont {Kazunori}\ \bibnamefont {Kohri}},
  \bibinfo {author} {\bibfnamefont {Yuuiti}\ \bibnamefont {Sendouda}}, \ and\
  \bibinfo {author} {\bibfnamefont {Jun'ichi}\ \bibnamefont {Yokoyama}},\
  }\bibfield  {title} {\enquote {\bibinfo {title} {{Constraints on primordial
  black holes}},}\ }\href {\doibase 10.1088/1361-6633/ac1e31} {\bibfield
  {journal} {\bibinfo  {journal} {Rept. Prog. Phys.}\ }\textbf {\bibinfo
  {volume} {84}},\ \bibinfo {pages} {116902} (\bibinfo {year} {2021})},\
  \Eprint {http://arxiv.org/abs/2002.12778} {arXiv:2002.12778 [astro-ph.CO]}
  \BibitemShut {NoStop}%
\bibitem [{\citenamefont {Kitajima}\ \emph {et~al.}(2023)\citenamefont
  {Kitajima}, \citenamefont {Lee}, \citenamefont {Murai}, \citenamefont
  {Takahashi},\ and\ \citenamefont {Yin}}]{Kitajima:2023cek}%
  \BibitemOpen
  \bibfield  {author} {\bibinfo {author} {\bibfnamefont {Naoya}\ \bibnamefont
  {Kitajima}}, \bibinfo {author} {\bibfnamefont {Junseok}\ \bibnamefont {Lee}},
  \bibinfo {author} {\bibfnamefont {Kai}\ \bibnamefont {Murai}}, \bibinfo
  {author} {\bibfnamefont {Fuminobu}\ \bibnamefont {Takahashi}}, \ and\
  \bibinfo {author} {\bibfnamefont {Wen}\ \bibnamefont {Yin}},\ }\bibfield
  {title} {\enquote {\bibinfo {title} {{Nanohertz Gravitational Waves from
  Axion Domain Walls Coupled to QCD}},}\ }\href@noop {} {\  (\bibinfo {year}
  {2023})},\ \Eprint {http://arxiv.org/abs/2306.17146} {arXiv:2306.17146
  [hep-ph]} \BibitemShut {NoStop}%
\bibitem [{\citenamefont {Guo}\ \emph {et~al.}(2023)\citenamefont {Guo},
  \citenamefont {Khlopov}, \citenamefont {Liu}, \citenamefont {Wu},
  \citenamefont {Wu},\ and\ \citenamefont {Zhu}}]{Guo:2023hyp}%
  \BibitemOpen
  \bibfield  {author} {\bibinfo {author} {\bibfnamefont {Shu-Yuan}\
  \bibnamefont {Guo}}, \bibinfo {author} {\bibfnamefont {Maxim}\ \bibnamefont
  {Khlopov}}, \bibinfo {author} {\bibfnamefont {Xuewen}\ \bibnamefont {Liu}},
  \bibinfo {author} {\bibfnamefont {Lei}\ \bibnamefont {Wu}}, \bibinfo {author}
  {\bibfnamefont {Yongcheng}\ \bibnamefont {Wu}}, \ and\ \bibinfo {author}
  {\bibfnamefont {Bin}\ \bibnamefont {Zhu}},\ }\bibfield  {title} {\enquote
  {\bibinfo {title} {{Footprints of Axion-Like Particle in Pulsar Timing Array
  Data and JWST Observations}},}\ }\href@noop {} {\  (\bibinfo {year}
  {2023})},\ \Eprint {http://arxiv.org/abs/2306.17022} {arXiv:2306.17022
  [hep-ph]} \BibitemShut {NoStop}%
\bibitem [{\citenamefont {Gouttenoire}\ and\ \citenamefont
  {Vitagliano}(2023)}]{Gouttenoire:2023ftk}%
  \BibitemOpen
  \bibfield  {author} {\bibinfo {author} {\bibfnamefont {Yann}\ \bibnamefont
  {Gouttenoire}}\ and\ \bibinfo {author} {\bibfnamefont {Edoardo}\ \bibnamefont
  {Vitagliano}},\ }\bibfield  {title} {\enquote {\bibinfo {title} {{Domain wall
  interpretation of the PTA signal confronting black hole overproduction}},}\
  }\href@noop {} {\  (\bibinfo {year} {2023})},\ \Eprint
  {http://arxiv.org/abs/2306.17841} {arXiv:2306.17841 [gr-qc]} \BibitemShut
  {NoStop}%
\bibitem [{\citenamefont {Unal}\ \emph {et~al.}(2023)\citenamefont {Unal},
  \citenamefont {Papageorgiou},\ and\ \citenamefont {Obata}}]{Unal:2023srk}%
  \BibitemOpen
  \bibfield  {author} {\bibinfo {author} {\bibfnamefont {Caner}\ \bibnamefont
  {Unal}}, \bibinfo {author} {\bibfnamefont {Alexandros}\ \bibnamefont
  {Papageorgiou}}, \ and\ \bibinfo {author} {\bibfnamefont {Ippei}\
  \bibnamefont {Obata}},\ }\bibfield  {title} {\enquote {\bibinfo {title}
  {{Axion-Gauge Dynamics During Inflation as the Origin of Pulsar Timing Array
  Signals and Primordial Black Holes}},}\ }\href@noop {} {\  (\bibinfo {year}
  {2023})},\ \Eprint {http://arxiv.org/abs/2307.02322} {arXiv:2307.02322
  [astro-ph.CO]} \BibitemShut {NoStop}%
\bibitem [{\citenamefont {Gouttenoire}\ \emph {et~al.}(2023)\citenamefont
  {Gouttenoire}, \citenamefont {Trifinopoulos}, \citenamefont {Valogiannis},\
  and\ \citenamefont {Vanvlasselaer}}]{Gouttenoire:2023nzr}%
  \BibitemOpen
  \bibfield  {author} {\bibinfo {author} {\bibfnamefont {Yann}\ \bibnamefont
  {Gouttenoire}}, \bibinfo {author} {\bibfnamefont {Sokratis}\ \bibnamefont
  {Trifinopoulos}}, \bibinfo {author} {\bibfnamefont {Georgios}\ \bibnamefont
  {Valogiannis}}, \ and\ \bibinfo {author} {\bibfnamefont {Miguel}\
  \bibnamefont {Vanvlasselaer}},\ }\bibfield  {title} {\enquote {\bibinfo
  {title} {{Scrutinizing the Primordial Black Holes Interpretation of PTA
  Gravitational Waves and JWST Early Galaxies}},}\ }\href@noop {} {\  (\bibinfo
  {year} {2023})},\ \Eprint {http://arxiv.org/abs/2307.01457} {arXiv:2307.01457
  [astro-ph.CO]} \BibitemShut {NoStop}%
\bibitem [{\citenamefont {Huang}\ \emph {et~al.}(2023)\citenamefont {Huang},
  \citenamefont {Cai}, \citenamefont {Jiang}, \citenamefont {Zhang},\ and\
  \citenamefont {Piao}}]{Huang:2023chx}%
  \BibitemOpen
  \bibfield  {author} {\bibinfo {author} {\bibfnamefont {Hai-Long}\
  \bibnamefont {Huang}}, \bibinfo {author} {\bibfnamefont {Yong}\ \bibnamefont
  {Cai}}, \bibinfo {author} {\bibfnamefont {Jun-Qian}\ \bibnamefont {Jiang}},
  \bibinfo {author} {\bibfnamefont {Jun}\ \bibnamefont {Zhang}}, \ and\
  \bibinfo {author} {\bibfnamefont {Yun-Song}\ \bibnamefont {Piao}},\
  }\bibfield  {title} {\enquote {\bibinfo {title} {{Supermassive primordial
  black holes in multiverse: for nano-Hertz gravitational wave and
  high-redshift JWST galaxies}},}\ }\href@noop {} {\  (\bibinfo {year}
  {2023})},\ \Eprint {http://arxiv.org/abs/2306.17577} {arXiv:2306.17577
  [gr-qc]} \BibitemShut {NoStop}%
\bibitem [{\citenamefont {Depta}\ \emph {et~al.}(2023)\citenamefont {Depta},
  \citenamefont {Schmidt-Hoberg},\ and\ \citenamefont
  {Tasillo}}]{Depta:2023qst}%
  \BibitemOpen
  \bibfield  {author} {\bibinfo {author} {\bibfnamefont {Paul~Frederik}\
  \bibnamefont {Depta}}, \bibinfo {author} {\bibfnamefont {Kai}\ \bibnamefont
  {Schmidt-Hoberg}}, \ and\ \bibinfo {author} {\bibfnamefont {Carlo}\
  \bibnamefont {Tasillo}},\ }\bibfield  {title} {\enquote {\bibinfo {title}
  {{Do pulsar timing arrays observe merging primordial black holes?}}}\
  }\href@noop {} {\  (\bibinfo {year} {2023})},\ \Eprint
  {http://arxiv.org/abs/2306.17836} {arXiv:2306.17836 [astro-ph.CO]}
  \BibitemShut {NoStop}%
\bibitem [{\citenamefont {Han}\ \emph {et~al.}(2023)\citenamefont {Han},
  \citenamefont {Xie}, \citenamefont {Yang},\ and\ \citenamefont
  {Zhang}}]{Han:2023olf}%
  \BibitemOpen
  \bibfield  {author} {\bibinfo {author} {\bibfnamefont {Chengcheng}\
  \bibnamefont {Han}}, \bibinfo {author} {\bibfnamefont {Ke-Pan}\ \bibnamefont
  {Xie}}, \bibinfo {author} {\bibfnamefont {Jin~Min}\ \bibnamefont {Yang}}, \
  and\ \bibinfo {author} {\bibfnamefont {Mengchao}\ \bibnamefont {Zhang}},\
  }\bibfield  {title} {\enquote {\bibinfo {title} {{Self-interacting dark
  matter implied by nano-Hertz gravitational waves}},}\ }\href@noop {} {\
  (\bibinfo {year} {2023})},\ \Eprint {http://arxiv.org/abs/2306.16966}
  {arXiv:2306.16966 [hep-ph]} \BibitemShut {NoStop}%
\bibitem [{\citenamefont {Choudhury}\ \emph {et~al.}(2023)\citenamefont
  {Choudhury}, \citenamefont {Panda},\ and\ \citenamefont
  {Sami}}]{Choudhury:2023rks}%
  \BibitemOpen
  \bibfield  {author} {\bibinfo {author} {\bibfnamefont {Sayantan}\
  \bibnamefont {Choudhury}}, \bibinfo {author} {\bibfnamefont {Sudhakar}\
  \bibnamefont {Panda}}, \ and\ \bibinfo {author} {\bibfnamefont
  {M.}~\bibnamefont {Sami}},\ }\bibfield  {title} {\enquote {\bibinfo {title}
  {{Quantum loop effects on the power spectrum and constraints on primordial
  black holes}},}\ }\href@noop {} {\  (\bibinfo {year} {2023})},\ \Eprint
  {http://arxiv.org/abs/2303.06066} {arXiv:2303.06066 [astro-ph.CO]}
  \BibitemShut {NoStop}%
\bibitem [{\citenamefont {{Batygin}}\ and\ \citenamefont
  {{Brown}}(2016)}]{2016AJ....151...22B}%
  \BibitemOpen
  \bibfield  {author} {\bibinfo {author} {\bibfnamefont {Konstantin}\
  \bibnamefont {{Batygin}}}\ and\ \bibinfo {author} {\bibfnamefont
  {Michael~E.}\ \bibnamefont {{Brown}}},\ }\bibfield  {title} {\enquote
  {\bibinfo {title} {{Evidence for a Distant Giant Planet in the Solar
  System}},}\ }\href {\doibase 10.3847/0004-6256/151/2/22} {\bibfield
  {journal} {\bibinfo  {journal} {Astron. J.}\ }\textbf {\bibinfo {volume}
  {151}},\ \bibinfo {eid} {22} (\bibinfo {year} {2016})},\ \Eprint
  {http://arxiv.org/abs/1601.05438} {arXiv:1601.05438 [astro-ph.EP]}
  \BibitemShut {NoStop}%
\bibitem [{\citenamefont {{Batygin}}\ \emph {et~al.}(2019)\citenamefont
  {{Batygin}}, \citenamefont {{Adams}}, \citenamefont {{Brown}},\ and\
  \citenamefont {{Becker}}}]{2019PhR...805....1B}%
  \BibitemOpen
  \bibfield  {author} {\bibinfo {author} {\bibfnamefont {Konstantin}\
  \bibnamefont {{Batygin}}}, \bibinfo {author} {\bibfnamefont {Fred~C.}\
  \bibnamefont {{Adams}}}, \bibinfo {author} {\bibfnamefont {Michael~E.}\
  \bibnamefont {{Brown}}}, \ and\ \bibinfo {author} {\bibfnamefont
  {Juliette~C.}\ \bibnamefont {{Becker}}},\ }\bibfield  {title} {\enquote
  {\bibinfo {title} {{The planet nine hypothesis}},}\ }\href {\doibase
  10.1016/j.physrep.2019.01.009} {\bibfield  {journal} {\bibinfo  {journal}
  {Phys. Rept.}\ }\textbf {\bibinfo {volume} {805}},\ \bibinfo {pages} {1--53}
  (\bibinfo {year} {2019})},\ \Eprint {http://arxiv.org/abs/1902.10103}
  {arXiv:1902.10103 [astro-ph.EP]} \BibitemShut {NoStop}%
\bibitem [{\citenamefont {{Trujillo}}\ and\ \citenamefont
  {{Sheppard}}(2014)}]{2014Natur.507..471T}%
  \BibitemOpen
  \bibfield  {author} {\bibinfo {author} {\bibfnamefont {Chadwick~A.}\
  \bibnamefont {{Trujillo}}}\ and\ \bibinfo {author} {\bibfnamefont {Scott~S.}\
  \bibnamefont {{Sheppard}}},\ }\bibfield  {title} {\enquote {\bibinfo {title}
  {{A Sedna-like body with a perihelion of 80 astronomical units}},}\ }\href
  {\doibase 10.1038/nature13156} {\bibfield  {journal} {\bibinfo  {journal}
  {\nat}\ }\textbf {\bibinfo {volume} {507}},\ \bibinfo {pages} {471--474}
  (\bibinfo {year} {2014})}\BibitemShut {NoStop}%
\bibitem [{\citenamefont {Scholtz}\ and\ \citenamefont
  {Unwin}(2020)}]{Scholtz:2019csj}%
  \BibitemOpen
  \bibfield  {author} {\bibinfo {author} {\bibfnamefont {Jakub}\ \bibnamefont
  {Scholtz}}\ and\ \bibinfo {author} {\bibfnamefont {James}\ \bibnamefont
  {Unwin}},\ }\bibfield  {title} {\enquote {\bibinfo {title} {{What if Planet 9
  is a Primordial Black Hole?}}}\ }\href {\doibase
  10.1103/PhysRevLett.125.051103} {\bibfield  {journal} {\bibinfo  {journal}
  {Phys. Rev. Lett.}\ }\textbf {\bibinfo {volume} {125}},\ \bibinfo {pages}
  {051103} (\bibinfo {year} {2020})},\ \Eprint
  {http://arxiv.org/abs/1909.11090} {arXiv:1909.11090 [hep-ph]} \BibitemShut
  {NoStop}%
\bibitem [{\citenamefont {{Mr{\'o}z}}\ \emph {et~al.}(2017)\citenamefont
  {{Mr{\'o}z}}, \citenamefont {{Udalski}}, \citenamefont {{Skowron}},
  \citenamefont {{Poleski}}, \citenamefont {{Koz{\l}owski}}, \citenamefont
  {{Szyma{\'n}ski}}, \citenamefont {{Soszy{\'n}ski}}, \citenamefont
  {{Wyrzykowski}}, \citenamefont {{Pietrukowicz}}, \citenamefont {{Ulaczyk}},
  \citenamefont {{Skowron}},\ and\ \citenamefont
  {{Pawlak}}}]{2017Natur.548..183M}%
  \BibitemOpen
  \bibfield  {author} {\bibinfo {author} {\bibfnamefont {Przemek}\ \bibnamefont
  {{Mr{\'o}z}}}, \bibinfo {author} {\bibfnamefont {Andrzej}\ \bibnamefont
  {{Udalski}}}, \bibinfo {author} {\bibfnamefont {Jan}\ \bibnamefont
  {{Skowron}}}, \bibinfo {author} {\bibfnamefont {Rados{\l}aw}\ \bibnamefont
  {{Poleski}}}, \bibinfo {author} {\bibfnamefont {Szymon}\ \bibnamefont
  {{Koz{\l}owski}}}, \bibinfo {author} {\bibfnamefont {Micha{\l}~K.}\
  \bibnamefont {{Szyma{\'n}ski}}}, \bibinfo {author} {\bibfnamefont {Igor}\
  \bibnamefont {{Soszy{\'n}ski}}}, \bibinfo {author} {\bibfnamefont
  {{\L}ukasz}\ \bibnamefont {{Wyrzykowski}}}, \bibinfo {author} {\bibfnamefont
  {Pawe{\l}}\ \bibnamefont {{Pietrukowicz}}}, \bibinfo {author} {\bibfnamefont
  {Krzysztof}\ \bibnamefont {{Ulaczyk}}}, \bibinfo {author} {\bibfnamefont
  {Dorota}\ \bibnamefont {{Skowron}}}, \ and\ \bibinfo {author} {\bibfnamefont
  {Micha{\l}}\ \bibnamefont {{Pawlak}}},\ }\bibfield  {title} {\enquote
  {\bibinfo {title} {{No large population of unbound or wide-orbit Jupiter-mass
  planets}},}\ }\href {\doibase 10.1038/nature23276} {\bibfield  {journal}
  {\bibinfo  {journal} {\nat}\ }\textbf {\bibinfo {volume} {548}},\ \bibinfo
  {pages} {183--186} (\bibinfo {year} {2017})},\ \Eprint
  {http://arxiv.org/abs/1707.07634} {arXiv:1707.07634 [astro-ph.EP]}
  \BibitemShut {NoStop}%
\bibitem [{\citenamefont {Niikura}\ \emph {et~al.}(2019)\citenamefont
  {Niikura}, \citenamefont {Takada}, \citenamefont {Yokoyama}, \citenamefont
  {Sumi},\ and\ \citenamefont {Masaki}}]{Niikura:2019kqi}%
  \BibitemOpen
  \bibfield  {author} {\bibinfo {author} {\bibfnamefont {Hiroko}\ \bibnamefont
  {Niikura}}, \bibinfo {author} {\bibfnamefont {Masahiro}\ \bibnamefont
  {Takada}}, \bibinfo {author} {\bibfnamefont {Shuichiro}\ \bibnamefont
  {Yokoyama}}, \bibinfo {author} {\bibfnamefont {Takahiro}\ \bibnamefont
  {Sumi}}, \ and\ \bibinfo {author} {\bibfnamefont {Shogo}\ \bibnamefont
  {Masaki}},\ }\bibfield  {title} {\enquote {\bibinfo {title} {{Constraints on
  Earth-mass primordial black holes from OGLE 5-year microlensing events}},}\
  }\href {\doibase 10.1103/PhysRevD.99.083503} {\bibfield  {journal} {\bibinfo
  {journal} {Phys. Rev. D}\ }\textbf {\bibinfo {volume} {99}},\ \bibinfo
  {pages} {083503} (\bibinfo {year} {2019})},\ \Eprint
  {http://arxiv.org/abs/1901.07120} {arXiv:1901.07120 [astro-ph.CO]}
  \BibitemShut {NoStop}%
\bibitem [{\citenamefont {Witten}(2020)}]{Witten:2020ifl}%
  \BibitemOpen
  \bibfield  {author} {\bibinfo {author} {\bibfnamefont {Edward}\ \bibnamefont
  {Witten}},\ }\bibfield  {title} {\enquote {\bibinfo {title} {{Searching for a
  Black Hole in the Outer Solar System}},}\ }\href@noop {} {\  (\bibinfo {year}
  {2020})},\ \Eprint {http://arxiv.org/abs/2004.14192} {arXiv:2004.14192
  [astro-ph.EP]} \BibitemShut {NoStop}%
\bibitem [{\citenamefont {Inomata}(2021)}]{Inomata:2020cck}%
  \BibitemOpen
  \bibfield  {author} {\bibinfo {author} {\bibfnamefont {Keisuke}\ \bibnamefont
  {Inomata}},\ }\bibfield  {title} {\enquote {\bibinfo {title} {{Analytic
  solutions of scalar perturbations induced by scalar perturbations}},}\ }\href
  {\doibase 10.1088/1475-7516/2021/03/013} {\bibfield  {journal} {\bibinfo
  {journal} {JCAP}\ }\textbf {\bibinfo {volume} {03}},\ \bibinfo {pages} {013}
  (\bibinfo {year} {2021})},\ \Eprint {http://arxiv.org/abs/2008.12300}
  {arXiv:2008.12300 [gr-qc]} \BibitemShut {NoStop}%
\bibitem [{\citenamefont {Dodelson}(2003)}]{Dodelson:2003ft}%
  \BibitemOpen
  \bibfield  {author} {\bibinfo {author} {\bibfnamefont {Scott}\ \bibnamefont
  {Dodelson}},\ }\href@noop {} {\emph {\bibinfo {title} {{Modern Cosmology}}}}\
  (\bibinfo  {publisher} {Academic Press},\ \bibinfo {address} {Amsterdam},\
  \bibinfo {year} {2003})\BibitemShut {NoStop}%
\end{thebibliography}%

\end{document}